\documentclass[11pt]{article}
\usepackage[margin=1in]{geometry}
\usepackage{mathtools,amssymb,amsthm}
\usepackage{tikz}
\usetikzlibrary{positioning}
\usepackage{subcaption}
\usepackage{natbib}
\usepackage{microtype}
\usepackage{aliascnt}
\usepackage[hidelinks,hypertexnames=false]{hyperref}
\allowdisplaybreaks
\newtheorem{theorem}{Theorem}
\newtheorem{lemma}{Lemma}
\newtheorem{definition}{Definition}
\newtheorem{claim}{Claim}
\newtheorem{remark}{Remark}
\newenvironment{claimproof}{\begin{proof}[Claim Proof]}{\end{proof}}

\newcommand{\size}[1]{|#1|}
\newcommand{\ceil}[1]{\lceil#1\rceil}
\newcommand{\floor}[1]{\lfloor#1\rfloor}
\newcommand{\lrA}[1]{\left(#1\right)}
\def\G{\mathcal{G}}
\def\I{\mathcal{I}}
\def\O{\mathcal{O}}
\def\T{\mathcal{T}}
\def\OPT{\mathrm{opt}}
\def\poly{\text{poly}}
\newcommand{\enc}{\mathcal{L}}
\DeclareMathOperator{\divg}{div}

\title{A Constant-Factor Approximation for Capacitated-Cable Facility Location with Hard Facility Capacities}
\author{Jingyang Zhao\thanks{Kyung Hee University. Email: \texttt{jingyangzhao1020@gmail.com}.}
\and
Mingyu Xiao\thanks{University of Electronic Science and Technology of China. Email: \texttt{myxiao@gmail.com}.}}
\date{}

\begin{document}
\maketitle
\begin{abstract}
Capacitated-cable facility location (CCFL) combines facility location and network design. Given clients with demands and candidate facilities with opening costs, the problem selects facilities and installs any number of cables of a common capacity on each edge to serve all demand at minimum total opening and cable cost. In the original problem, facilities have unlimited capacity.
We study the variant with hard facility capacities (H-CCFL), in which each facility can open at most once and serve at most its capacity. For split demands, where a client's demand may be served by multiple facilities, we obtain the first constant-factor approximation algorithm on general metrics.
As applications, we also obtain the first constant-factor approximation algorithms on general metrics for several well-known problems, including the airport and railway problem with arbitrary opening costs and the multiple-depot split delivery vehicle routing problem with limited fleets.
\end{abstract}

\section{Introduction}\label{sec:intro}
A company serving clients at different locations must decide where to open facilities and how to transport goods between the facilities and the clients. These facility location (FL) and network design (ND) decisions interact: facility locations affect the transportation network, while shipments for different clients can share capacity on the same links \citep{journal:meyerson:SICOMP:08,journal:contreras:EJOR:12}. For example, when transportation is charged by the number of trucks used, a partially loaded truck can cost as much as a full one. Combining shipments on common links can therefore reduce transportation costs \citep{journal:ravi:OR:06,journal:arulselvan:IJOC:17}.
Related decisions also arise in telecommunications and power distribution, where facilities such as switches or substations must be connected to clients and both facilities and links may have limited capacities \citep{journal:melkote:EJOR:01,journal:schwarze:CEJOR:21}. In airport and railway design, airports and railway connections are also chosen together to serve cities \citep{conf:adamaszek:STACS:16}.

When facilities have unlimited capacity, \citet{journal:ravi:OR:06} modeled this setting as the capacitated-cable FL (CCFL). An input instance is $\I_{\mathrm{CCFL}}=(G,c,J,U,d,\phi,Q)$. The graph $G=(V,E)$ is an undirected complete graph with metric edge costs $c:E\to\mathbb{Q}_{\ge0}$. We use $c(u,v)$ for the cost of edge $uv$ and set $c(v,v)=0$. Each node represents a location. The sets $J,U\subseteq V$ contain the clients and candidate facilities, respectively. A node may contain both a client and a facility. Each client $v\in J$ has demand $d_v\in\mathbb{N}^+$, and $d_v=0$ for $v\notin J$. Each facility $r\in U$ has opening cost $\phi_r\in\mathbb{Q}_{\ge0}$. The vectors $d$ and $\phi$ collect these demands and opening costs. A cable on edge $e$ costs $c_e=c(e)$ and carries at most $Q\in\mathbb{N}^+$ units.
A solution opens a set $P\subseteq U$ of facilities and installs $k_e\in\mathbb{Z}_{\ge0}$ cables on each edge $e$. The demand is sent from clients to open facilities along the installed cables, with at most $Qk_e$ units on each edge $e$. With split demands, each client may be served by multiple facilities; with unsplit demands, each client must be served by one facility. Demand may split among paths, and any node may relay it; in particular, nodes outside $J\cup U$ may serve as Steiner nodes. A client at an open facility may be served there without using a cable, and each facility has no capacity limit. The objective is to minimize the total opening and cable cost $\sum_{r\in P}\phi_r+\sum_{e\in E}c_ek_e$.

Let $\rho_{\mathrm{UFL}}$ and $\rho_{\mathrm{ST}}$ denote approximation ratios for the uncapacitated FL (UFL) and the Steiner tree (ST), respectively.
\citet{journal:ravi:OR:06} obtained a $(\rho_{\mathrm{UFL}}+\rho_{\mathrm{ST}})$-approximation for split demands with $d_v=1$, and a $(2\rho_{\mathrm{UFL}}+\rho_{\mathrm{ST}})$-approximation for unsplit demands with $d_v\le Q$.

To account for facility capacities, \citet{journal:chen:ALG:09} extended the CCFL to allow soft facility capacities. Each facility $r$ has capacity $b_r\in\mathbb{N}^+$, and any number of copies may open at the same location. Each copy incurs the opening cost $\phi_r$ and can serve at most $b_r$ units of demand. They obtained a $19.84$-approximation for split demands and a $28.06$-approximation for unsplit demands, where a client may be served by one or more copies at a single location. They also developed constant-factor approximation algorithms for variants requiring a connected cable network or a separate network for communication among open facilities.

In this paper, we study the CCFL with hard facility capacities (H-CCFL), a variant of the CCFL in which each facility $r\in U$ can open at most once and serve at most $b_r$ units of demand. This problem was posed as an open question by \citet{journal:chen:ALG:09}. We focus on split demands.

Hard capacities arise in distribution networks, where a facility may have a limited capacity or a limited number of vehicles \citep{journal:baldacci:OR:11,journal:gouveia:TS:23}. Unlike soft capacities, these limits cannot be increased by opening additional copies at the same location.
In the H-CCFL, clients may share cables, and a cable costs the same whether it carries one unit or $Q$ units. Connecting each client separately can therefore leave cables partially filled and increase the total cost. At the same time, a low-cost cable network may not reach enough facility capacity to serve all clients. Thus, the challenge is to select facilities with enough capacity and connect all clients at low cost using shared cables.
Hard capacities commonly arise in ND and vehicle routing problems (VRPs). We next describe three related problems that feature hard capacities.

\begin{enumerate}
\item The airport and railway problem (ARP) models the joint design of airports and railway connections. Given a complete graph on a set $V$ of cities with metric edge costs, an airport may open at each city at a specified cost. A solution partitions the cities into trees, each with exactly one open airport and at most $K$ cities, including its airport. The objective is to minimize the total tree and opening cost \citep{conf:adamaszek:STACS:16}. For arbitrary opening costs on general metrics, the best previously known approximation was an $\O(\log\size{V})$-approximation based on randomized tree embeddings \citep{journal:salavatipour:JOCO:25,journal:fakcharoenphol:JCSS:04}.

\item The multiple-depot split delivery VRP (MDSDVRP) models distribution from multiple depots \citep{book:toth:SIAM:14}.
Given clients and depots with metric travel costs, each client has a demand, and each depot $r$ has $m_r$ vehicles of capacity $Q$.
A client's demand may be split among vehicles from different depots, which can reduce delivery costs \citep{journal:dror:TS:89}.
Each vehicle serves at most one route, which starts and ends at its depot, so depot $r$ can deliver at most $m_rQ$ units in total.
The objective is to minimize the total route cost.
The MDSDVRP is well known, and both exact and heuristic algorithms have been widely studied \citep{journal:gouveia:TS:23}. With an unlimited fleet at each depot, a constant-factor approximation has long been known \citep{journal:li:ORSAC:90}. However, for limited fleets, previous constant-factor algorithms had running times parameterized by the number of depots \citep{journal:lai:IJOC:23,preprint:zhao:arxiv:26a}.

\item The capacitated location routing (CLR) combines depot selection with vehicle routing in distribution networks \citep{journal:baldacci:OR:11,journal:contardo:IJOC:14}. In the CLR with hard facility capacities (H-CLR), we are given clients and candidate depots with metric travel costs. Each client has a demand, and each depot has an opening cost and a hard capacity on the total demand it can deliver. Given vehicle capacity $Q$, each route starts and ends at an open depot and carries at most $Q$ units. Demand may split among routes and depots, and the objective is to minimize the total depot opening and route cost. To our knowledge, the only previously known approximation result for general depot capacities was a bifactor approximation algorithm that may violate depot capacities \citep{journal:carrasco:EJOR:23}.
\end{enumerate}

We formally define these problems and review related approximation algorithms in Section~\ref{sec:applications}.

\subsection{Our Results}\label{sec:results}
Our main result is a polynomial-time $\Gamma$-approximation for the H-CCFL with split demands, where $\Gamma=2^{33}$. (We did not attempt to optimize this constant.) As applications, we reduce the ARP, the MDSDVRP, and the H-CLR to the H-CCFL and convert the resulting cable solutions into trees or vehicle routes. This yields polynomial-time approximations with ratios $3\Gamma$ for the ARP, $6\Gamma$ for the MDSDVRP, and $20\Gamma$ for the H-CLR.
For the H-CLR, we require each depot capacity to be at least the vehicle capacity, a natural assumption when each depot can supply at least a fully loaded vehicle.
For instance, this assumption holds in benchmark instances from the Prins \citep{journal:prins:FOUROR:06}, Barreto \citep{journal:barreto:EJOR:07}, and Akca \citep{conf:akca:ICS:09} sets, as well as in the Schneider set \citep{journal:schneider:TS:19}.
Thus, we obtain the first constant-factor approximations on general metrics for the H-CCFL, the ARP with arbitrary opening costs, and the MDSDVRP, as well as for the H-CLR when each depot capacity is at least the vehicle capacity.

We focus on the H-CCFL with split demands. In contrast, for unsplit demands, a reduction from bin packing shows that no polynomial-time $\poly(\size{V})$-approximation exists unless $\mathrm{P}=\mathrm{NP}$, even when $d_v\le Q$ for every client $v$. For an instance with $n$ items of size at most $Q$ and $k$ bins of capacity $Q$, create a client for each item with demand equal to its size, $k$ facilities of capacity $Q$ and opening cost $1$, and one facility with capacity equal to the total demand and opening cost $M$. Place them at distinct nodes, with edge cost $1$ between every pair. If the items fit in the $k$ bins, a solution costs at most $n+k$; otherwise, every feasible solution opens the facility of cost $M$.
For any fixed polynomial $\poly(\size{V})$, we can choose an integer $M>\poly(\size{V})(n+k)$, ruling out a polynomial-time $\poly(\size{V})$-approximation.

\subsection{Related Work}\label{sec:related}
Unless stated otherwise, all problems considered in this paper are on general metrics.

\paragraph{Capacitated facility location.}
The CCFL and the H-CCFL are closely related to FL. When $Q=1$, the CCFL becomes the UFL, and the H-CCFL with split demands is the capacitated FL (CFL): we assume that the CFL has hard facility capacities, where each facility may open at most once and serve at most its capacity.
For the CFL with split demands and nonuniform facility capacities, \citet{conf:pal:FOCS:01} developed the first constant-factor approximation, with ratio $9+\varepsilon$, and \citet{conf:bansal:ESA:12} later improved the ratio to $5+\varepsilon$, for any fixed $\varepsilon>0$.
For the standard fractional relaxation of the CFL with unit demands, \citet{report:abrams:CMU:02} proposed a rounding algorithm with cost at most $18$ times the fractional cost, while each facility may receive at most twice its capacity.
This result also applies to general split demands \citep{journal:an:SICOMP:17}.
For an overview of approximations for FL and its variants, see \citet{book:williamson:CUP:11} and \citet{journal:xiao:MATHEMATICS:25}.

\paragraph{Multiple cable types.}
The CCFL with $K$ cable types (CCFL-$K$) extends the CCFL by allowing multiple cable types. Each type has a capacity and a cost factor; installing a copy on an edge costs the factor times the edge cost. Any number of copies of each type may be installed on an edge, and facilities have no capacity limits.
For the CCFL-$K$ with unit demands, \citet{conf:ravi:IPCO:02} obtained an $\O(K)$-approximation. When cable types satisfy \emph{economies of scale} (larger cables have higher cost factors but lower costs per unit of capacity) and each client's demand is routed along a single path to a facility, \citet{journal:friggstad:ALG:19} obtained an $\O(K)$-approximation using linear programming (LP) rounding, and an $\O(1)$-approximation for a connected variant in which a separate network of unlimited-capacity cables connects the open facilities.

The buy-at-bulk FL (BBFL) further generalizes the CCFL-$K$: each client may be served by multiple facilities, and routing $a$ units on edge $e$ costs $c_e g(a)$, where $g$ is a common nondecreasing and subadditive function satisfying $g(a+b)\le g(a)+g(b)$ for all $a,b\in\mathbb{Z}_{\ge0}$. On trees, \citet{conf:nematollahi:WAOA:26} developed a polynomial-time approximation scheme (PTAS) for the BBFL, provided that $g$ can be approximated within $1+\varepsilon$ in polynomial time for any fixed $\varepsilon>0$.

\paragraph{Buy-at-bulk network design.}
The buy-at-bulk ND (BBND) seeks a minimum-cost network that routes demands between $h$ given pairs of nodes. Related models include capacitated telecommunications network design and variants with protection against network failures \citep{journal:magnanti:OR:95,journal:salman:SIOPT:01,journal:antonakopoulos:MOR:11}. Unlike the BBFL, the BBND does not select facilities.

In the uniform case, routing $a$ units on edge $e$ costs $c_e g(a)$, where every edge uses the same nondecreasing subadditive function $g$. When cable costs satisfy economies of scale, the CCFL-$K$ with one zero-cost facility is a single-sink case. For the single-sink case with the same cable types on every edge, \citet{journal:guha:SICOMP:09} gave a randomized $292$-approximation. For multiple demand pairs, an $\O(\log\size{V})$-approximation is known via randomized tree embeddings \citep{conf:awerbuch:FOCS:97,journal:fakcharoenphol:JCSS:04}. For this case, \citet{conf:andrews:FOCS:04} showed that no $\O(\log^{\frac{1}{4}-\varepsilon}\size{V})$-approximation exists unless $\mathrm{NP}\subseteq\mathrm{ZPTIME}(N^{\operatorname{polylog}N})$. Moreover, a special case is the rent-or-buy ND, where routing $a$ units on edge $e$ costs $c_e\min\{a,M\}$ for a parameter $M>0$: capacity can be rented at cost $c_e$ per unit or purchased without limit for $Mc_e$. For multiple demand pairs, \citet{journal:fleischer:SICOMP:10} obtained a randomized $5$-approximation.

In the nonuniform case, each edge $e$ and node $v$ may have its own nondecreasing concave cost function $g_e$ and $g_v$, respectively. If $x_e$ and $x_v$ are their total routed demands, \citet{journal:chekuri:SICOMP:10} developed an $\O(\log^4 h)$-approximation algorithm for cost $\sum_{e\in E}g_e(x_e)+\sum_{v\in V}g_v(x_v)$. They also noted that, with node costs, even the single-sink case admits no $o(\log h)$-approximation unless $\mathrm{P}=\mathrm{NP}$, by a reduction from set cover. Recently, \citet{preprint:gholami:arxiv:26} improved the approximation ratio to $\O(\frac{\log^2 h}{(\log\log h)^2})$.

\section{Preliminaries}\label{sec:tools}\label{sec:model}
Let $\I=(G,c,J,U,d,b,\phi,Q)$ denote an instance of the H-CCFL described in Section~\ref{sec:intro}. Here $G=(V,E)$ is a complete graph with metric edge costs $c$; $J$ and $U$ are the sets of clients and facilities; $d$, $b$, and $\phi$ are the vectors of demands, facility capacities, and opening costs; and $Q$ is the cable capacity.
Let $n=\size{J}$, $m=\size{U}$, and $d_J=\sum_{v\in J}d_v$. For a set $P$ of facilities, let $\Phi(P)=\sum_{r\in P}\phi_r$.
The instance is feasible exactly when $\sum_{r\in U}b_r\ge d_J$, so we may assume the instance is feasible.

Let $\OPT$ be the optimum of the instance under consideration. Demands and capacities are encoded in binary and may be superpolynomial in $\size{V}$.
Let $\enc$ be the input size of $\I$; all polynomial-time claims refer to $\enc$.
A zero-cost solution, if one exists, can be found in polynomial time by a maximum-flow computation using facilities of zero opening cost and zero-cost edges \citep{book:schrijver:springer:03}, so we may also assume $\OPT>0$.

We represent the routing of demand from clients to facilities as a flow $f$ on $G$. Choose an orientation for each edge, and let $f=(f_e)_{e\in E}$, where $f_e$ is the signed net flow in that direction. Let $\divg f(v)$ denote the net flow leaving node $v$: the sum of $f_e$ over edges oriented away from $v$ minus the sum over edges oriented toward $v$. We call the amount of demand assigned to a facility its \emph{receipt}; the receipt of a facility from a client is the amount of that client's demand assigned to it, and the receipt of a set of facilities is the total amount assigned to them. Let $q_r$ be the receipt of facility $r$ and set $q_v=0$ for $v\notin U$. Flow conservation and facility capacities require
\begin{equation}\label{eq:model}
 \divg f(v)=d_v-q_v\quad(v\in V),\qquad 0\le q_r\le b_r\quad(r\in U).
\end{equation}
We call a flow from clients to facilities \emph{integral} if $f_e\in\mathbb{Z}$ for every edge $e$ and $q_r\in\mathbb{Z}_{\ge0}$ for every facility $r$. To serve this flow using open facilities $P\subseteq U$, we require $q_r=0$ for each $r\notin P$ and must install at least $\ceil{\frac{|f_e|}{Q}}$ cables on each edge $e$. Thus, the minimum cable cost for $f$ is $C_Q(f)=\sum_e c_e\ceil{\frac{|f_e|}{Q}}$.
Since demands and capacities are integers, any feasible solution can be converted into one with an integral flow at no greater cost by integrality of maximum flow \citep{book:schrijver:springer:03}. We may cancel directed cycles without increasing cable cost. Every edge then carries at most $d_J$ units, so we may cap $b_r$ and $Q$ at $d_J$ whenever $d_J>0$.

We next introduce cost notation used throughout the paper.
For an edge set, tree, or path $H$, let $c(H)=\sum_{e\in H}c_e$, counting repeated edges with multiplicity. For a flow $f$, the edges carrying nonzero flow form its \emph{support}, and we let $L(f)=\sum_e c_e|f_e|$. Since $\ceil{\frac{|u+v|}{Q}}\le\ceil{\frac{|u|}{Q}}+\ceil{\frac{|v|}{Q}}$, the cable cost satisfies
\begin{equation}\label{eq:subadd}
 \frac{L(f)}{Q}\le C_Q(f),\qquad C_Q(f+g)\le C_Q(f)+C_Q(g).
\end{equation}

After serving part of the demand, we consider the \emph{residual instance}. In this instance, $J$, $d$, $b$, and $\phi$ refer to the current clients, demands, facility capacities, and opening costs; a facility's remaining capacity may be zero, and an already opened facility has zero opening cost. A client is removed from $J$ once its remaining demand is zero, while its node remains in $V$. A facility is \emph{free} if its current opening cost is zero and \emph{nonfree} otherwise.
We compare the residual instance with a solution for the original demand using the following lemma, which gives a flow without increasing any facility receipt. Let $(x)_+=\max\{x,0\}$.

\begin{lemma}\label{lem:residual-flow}
Let $f$ and $g$ be integral flows from clients to facilities with integral demand vectors $d^f$ and $d^g$, respectively, where $0\le d^g\le d^f$. Suppose facility $r$ receives integral amounts $q_r^f$ under $f$ and $q_r^g$ under $g$. There exists an integral flow $h$ that sends the remaining demand $d^f-d^g$ to facilities, with at most $(q_r^f-q_r^g)_+$ units received at each facility $r$ and edge loads $|h_e|\le|f_e-g_e|\le|f_e|+|g_e|$.
\end{lemma}
\begin{proof}
Let $\Delta=f-g$. Orient each edge according to the sign of $\Delta_e$, with flow $|\Delta_e|$. Each client $v$ supplies $d_v^f-d_v^g$, and each facility $r$ supplies $(q_r^g-q_r^f)_+$ or receives $(q_r^f-q_r^g)_+$. Add two auxiliary nodes $s$ and $t$: connect $s$ to each client and facility with arcs carrying their supplies, and connect each facility $r$ to $t$ with an arc carrying $(q_r^f-q_r^g)_+$, keeping these arcs separate even when a client and a facility share a node. Every original node then satisfies flow conservation.
By the flow decomposition theorem \citep{book:ahuja:prentice:93}, we can decompose the resulting flow into integral $s$--$t$ paths and cycles. Discard the cycles and paths starting with arcs for facility supplies. After removing the auxiliary arcs, the remaining paths form an integral flow $h$ serving $d^f-d^g$. Let $q_r^h$ denote the receipt of facility $r$ under $h$.
The retained paths deliver $q_r^h\le(q_r^f-q_r^g)_+$ units to each facility $r$ and carry $|h_e|\le|\Delta_e|=|f_e-g_e|\le|f_e|+|g_e|$ units on each original edge $e$.
\end{proof}

\section{Global-to-Local Reduction via Trees}\label{sec:global}
We reduce the H-CCFL to local instances with a given tree connecting their clients.
In particular, we use a candidate edge vector $x$ to partition the nodes into sets and form a rooted tree $\T^{\mathrm{reg}}$ whose nodes are these sets. The union of each set and its descendant sets defines a region. For each set, we also construct a tree connecting its nodes. We then process the regions from the leaves to the root of $\T^{\mathrm{reg}}$, using these trees as input to a local algorithm. We will show that a polynomial-time local algorithm whose cost is bounded by a constant times the cost of any feasible solution plus a constant times the cost of its input tree yields a polynomial-time constant-factor approximation for the H-CCFL.

Our reduction adapts the local-to-global idea of \citet{preprint:zhao:arxiv:26b} for covering disjointness-compliable set families. Their algorithm merges the node sets selected by each local tree without paying again for the connections already made. In the H-CCFL, however, sending additional demand along previously used edges may require additional cables, and facilities must still be treated individually according to their remaining capacities and opening costs. Therefore, after each local solution, we instead update the remaining demands, facility capacities, and opening costs.

To find a suitable edge vector $x$, we test candidate vectors by either constructing a solution or returning a violated inequality. The following lemma follows from the ellipsoid method \citep{book:grotschel:springer:93}.

\begin{lemma}[Separation or construction]\label{lem:constructive-feasibility}
Let $\mathcal P$ be a rational polytope of polynomial dimension that can be defined by inequalities whose individual encoding lengths have a known polynomial bound.
Suppose an oracle, given a rational point $x$, either constructs a desired solution or returns a polynomial-size inequality valid for $\mathcal P$ and violated by $x$. Then a desired solution can be found or $\mathcal P$ can be shown to be empty using polynomially many oracle calls and polynomial additional computation.
\end{lemma}

\subsection{Trees and Small Cuts}
Given a nonnegative edge vector $x$, we first partition $V$ into disjoint node sets $S_i$ and make each set a node of $\T^{\mathrm{reg}}$. For each node set $S_i$, its \emph{region} $V_i$ is the union of that set and all its descendant sets in $\T^{\mathrm{reg}}$. We also construct a tree $\T_i$ in $G$ within $S_i$ that connects all its nodes. When we process regions from the leaves to the root, every remaining client in $V_i$ lies in $S_i$ because the clients in its descendant sets have already been served, so $\T_i$ connects these clients.

For a node set $W\subseteq V$, let its \emph{boundary} $\partial W$ be the set of edges with exactly one endpoint in $W$. For any edge vector $x$, let $x(\partial W)=\sum_{e\in\partial W}x_e$ and $c(x)=\sum_e c_ex_e$. We seek regions whose boundaries have small $x$-values and trees with small total cost. The boundary bound helps us compare the local solutions with a global solution; the tree-cost bound controls the cost of the local solutions.

\begin{figure}[!t]
\centering
\def\HCCFLcomponents{
  \draw[thick] (0.30,2.30)--(0.60,2.00)--(0.90,2.30) (0.60,2.00)--(0.60,1.72);
  \draw[thick] (0.30,0.55)--(0.65,0.55)--(0.35,0.85)--(0.60,1.10);
  \draw[thick] (2.85,2.30)--(3.15,2.00)--(3.55,2.16) (3.15,2.00)--(3.15,1.72);
  \draw[thick] (2.85,0.55)--(3.35,0.55)--(3.35,0.90)--(3.15,1.10);
  \foreach \xx/\yy in {0.30/2.30,0.60/2.00,0.90/2.30,0.60/1.72,
    0.30/0.55,0.65/0.55,0.35/0.85,0.60/1.10,
    2.85/2.30,3.15/2.00,3.55/2.16,3.15/1.72,
    2.85/0.55,3.35/0.55,3.35/0.90,3.15/1.10}
    \fill (\xx,\yy) circle (1.1pt);
}
\begin{subfigure}[t]{0.32\textwidth}
\centering
\begin{tikzpicture}[x=0.9cm,y=0.9cm,every node/.style={font=\footnotesize}]
  \draw[blue!65!black,dashed,thick,rounded corners=3pt] (2.10,0.15) rectangle (4.20,2.75);
  \node[anchor=south west,inner sep=1pt] at (2.10,2.75) {$W_1$};
  \draw[gray,densely dotted,line width=0.9pt] (0.60,1.72)--(0.60,1.10) (3.15,1.72)--(3.15,1.10);
  \HCCFLcomponents
\end{tikzpicture}
\caption{Forest $F$.}
\end{subfigure}\hfill
\begin{subfigure}[t]{0.32\textwidth}
\centering
\begin{tikzpicture}[x=0.9cm,y=0.9cm,every node/.style={font=\footnotesize}]
  \draw[blue!65!black,dashed,thick,rounded corners=3pt] (2.10,0.15) rectangle (4.20,2.75);
  \node[anchor=south west,inner sep=1pt] at (2.10,2.75) {$W_1$};
  \draw[red!70!black,dashed,thick,rounded corners=3pt] (2.55,0.35) rectangle (3.75,1.35);
  \node[anchor=south east,inner sep=1pt] at (4.16,1.35) {$W_2$};
  \draw[red!70!black,dashed,thick,rounded corners=3pt] (0.10,0.35) rectangle (1.10,1.30);
  \node[anchor=south west,inner sep=1pt] at (0.10,1.30) {$W_3$};
  \HCCFLcomponents
  \node[anchor=west,inner sep=1pt] at (0.93,2.00) {\scriptsize $S_1$};
  \node[anchor=west,inner sep=1pt] at (0.66,0.83) {\scriptsize $S_4$};
  \node[anchor=west,inner sep=1pt] at (3.60,2.00) {\scriptsize $S_2$};
  \node[anchor=west,inner sep=1pt] at (3.39,1.12) {\scriptsize $S_3$};
\end{tikzpicture}
\caption{Node sets in $H$.}
\end{subfigure}\hfill
\begin{subfigure}[t]{0.32\textwidth}
\centering
\begin{tikzpicture}[
  x=0.9cm,y=0.9cm,every node/.style={font=\footnotesize},
  setnode/.style={draw,rounded corners=2pt,align=center,minimum width=1.28cm,minimum height=0.72cm,inner sep=2pt}
]
  \node[setnode] (root) at (1.85,2.65) {$S_1$\\[-1pt]$V_1=V$};
  \node[setnode] (middle) at (0.65,1.42) {$S_2$\\[-1pt]$V_2=W_1$};
  \node[setnode] (right) at (3.05,1.42) {$S_4$\\[-1pt]$V_4=W_3$};
  \node[setnode] (leaf) at (0.65,0.19) {$S_3$\\[-1pt]$V_3=W_2$};
  \draw[thick] (root)--(middle) (root)--(right) (middle)--(leaf);
\end{tikzpicture}
\caption{The rooted tree $\T^{\mathrm{reg}}$.}
\end{subfigure}
\caption{Selected sets and the rooted tree: blue from $F$, red from deletions; dotted edges are deleted.}
\label{fig:global-forest-partition}
\end{figure}
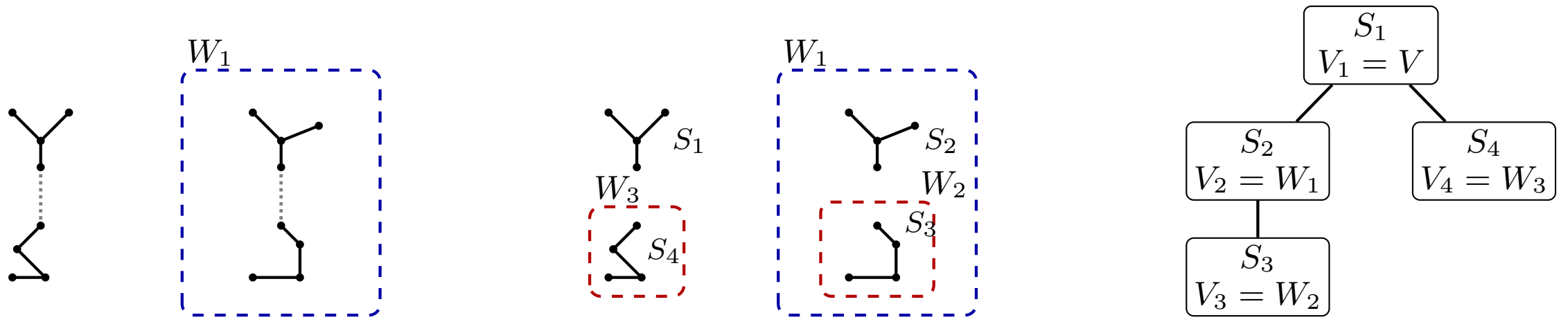

\begin{lemma}[Trees with small boundaries]\label{lem:trees-small-boundaries}
For a connected undirected graph with nonnegative rational edge costs, a nonnegative rational edge vector $x$, and a positive rational $\delta$, there exists a polynomial-time algorithm that partitions $V$ into node sets $S_i$, forms a rooted tree $\T^{\mathrm{reg}}$ whose nodes are these sets, and finds a spanning tree $\T_i$ within each set. Moreover, $x(\partial V_i)<\delta$ for every nonroot node $S_i$ of $\T^{\mathrm{reg}}$ and $\sum_i c(\T_i)\le\frac{2c(x)}{\delta}$.
\end{lemma}
\begin{proof}
We adapt the forest growth and edge-deletion method of \citet{journal:goemans:SICOMP:95}.

\paragraph{Primal--dual growth.}
We first grow a forest whose components have small boundaries. Start with a forest on $V$ with no edges, so each node is a component. For the construction, set temporary edge values $w_e=c_e$.
A current component $W$ is active if $x(\partial W)\ge\delta$; we need to merge it because its boundary is too large. We view the growth as a continuous process. For each $W\subseteq V$, let $y_W$ be the total time for which $W$ is an active component. While an active component exists, decrease each $w_e$ at a rate equal to the number of active component boundaries containing $e$. When an edge of positive rate reaches zero, add it to the forest and merge its components. If an edge between different components already has zero remaining value and meets an active component, add it immediately, using a fixed edge order to break ties.
Record each distinct inactive component when it appears, including inactive singletons and all final components.
We stop when all components are inactive and denote the forest by $F$. Since components only merge, any two recorded sets are disjoint or one contains the other.

\paragraph{Pruning.}
We first select all but one component of $F$; each is inactive. We then set $H=F$ and delete edges that are unnecessary for the final trees. Whenever a recorded inactive set has exactly one edge of $H$ across its boundary, we delete that edge and select the set. We continue until no such set remains. These selected sets are distinct: each deletion removes the only edge of $H$ crossing its selected set, and no component of $F$ has an edge of $H$ across its boundary. If $H$ has $k$ components, there are $k-1$ selected sets: we selected all but one component of $F$, and each deletion adds one component and one selected set.

\paragraph{Forming the node sets.}
We use the selected sets to form the node sets.
The selected sets separate any two components of $H$: components belonging to different trees of $F$ are separated by one of the initially selected sets; for two components belonging to the same tree of $F$, the set selected when the first deletion occurs on their path in $F$ separates them.
No selected boundary crosses a component of $H$.

Add $V$ as the root of a tree. For each selected set, its parent is the smallest set among the selected sets and $V$ that strictly contains it. Assign each node to the smallest set containing it. Let $V_i$ denote a set in this tree and $S_i$ the nodes assigned to it. Relabel each tree node $V_i$ by $S_i$ to obtain $\T^{\mathrm{reg}}$. Thus each node set consists of the nodes in its region but not in its child regions.
Since no selected boundary crosses a component of $H$ and the selected sets separate its components, each component is assigned to a different node set. There are $k$ components and $k$ node sets, so each node set is one component of $H$. Its edges in $H$ form a spanning tree $\T_i$. Each nonroot region is a selected inactive set, so its boundary has $x$-value less than $\delta$. An illustration of the selection, deletion, and resulting rooted tree is shown in Figure~\ref{fig:global-forest-partition}.

We next bound $c(H)$. Since $w_e$ remains nonnegative, $\sum_{W\subseteq V:e\in\partial W}y_W\le c_e$ for every edge $e$; equality holds for each edge of $F$. Each active component satisfies $x(\partial W)\ge\delta$, so
\[
 \delta\sum_{W\subseteq V} y_W\le\sum_{W\subseteq V} x(\partial W)y_W
 =\sum_e x_e\sum_{W\subseteq V:e\in\partial W}y_W\le c(x).
\]

Consider a growth interval with active components $\mathcal A$. Contract the components present in that interval (each is connected in F), and retain only the crossing edges of $H$. The result is a forest because each contracted component is connected in $F$. Every nonisolated inactive node has degree at least two: otherwise its recorded set would have one edge of $H$ across its boundary and the deletion step would have removed it. The degree sum of the forest therefore gives $\sum_{W\in\mathcal A}\size{H\cap\partial W}\le2\size{\mathcal A}$. Summing over the growth intervals and using the equality above for edges of $H$ yields
\[
 \sum_i c(\T_i)=c(H)=\sum_{W\subseteq V} y_W\size{H\cap\partial W}\le2\sum_{W\subseteq V} y_W\le\frac{2c(x)}{\delta}.
\]

There are at most $\size{V}-1$ merges and as many deletions, so the construction runs in polynomial time.
\end{proof}

\subsection{Global-to-Local Reduction}
The local subproblem is a feasible residual H-CCFL instance together with a tree $\T$ connecting its clients. We seek a feasible solution of minimum opening and cable cost under the current demands and facility capacities.

We process the regions from the leaves to the root. For region $V_i$, the local instance uses the induced subgraph $G[V_i]$ with its edge costs, the remaining demands and facility capacities in $V_i$, and the current opening costs (zero for facilities opened earlier), together with $\T_i$. Since $G$ is complete, $G[V_i]$ is a complete graph with metric costs. All remaining clients in $V_i$ lie in $S_i$ and are connected by $\T_i$. The following lemma shows how a local guarantee that also depends on the cost of this tree yields a global approximation.

\begin{lemma}[Global-to-local reduction]\label{lem:global}
Suppose a local algorithm $\varUpsilon$, given a feasible residual H-CCFL instance and a tree $\T$ connecting its clients, returns a feasible solution of cost at most $\alpha(\Phi+C)+\beta c(\T)$ for every feasible solution of that instance with opening cost $\Phi$ and cable cost $C$.
For fixed rational constants $\alpha\ge1$ and $\beta\ge0$, if $\varUpsilon$ runs in time $\poly(\enc)$, then the H-CCFL admits a $4[\alpha+4\beta(1+\alpha)]$-approximation in time $\poly(\enc)$.
\end{lemma}
\begin{proof}
Let $\kappa=\alpha+4\beta(1+\alpha)$ and $\delta=\frac{1}{2(1+\alpha)}$. We may assume $\OPT>0$. If $\size{V}=1$, one call to $\varUpsilon$ with the singleton tree gives a solution of cost at most $\alpha\OPT$, so we also assume $\size{E}>0$.

For a fixed budget $B>0$, we seek a feasible solution of cost at most $2\kappa B$. We first describe how to process the regions for a candidate edge vector $x$. We define a polytope $\mathcal Q_B$ using constraints on the edge values and the total cost of processing the regions. We then show that $\mathcal Q_B$ contains, for every feasible integral solution of cost at most $B$, the vector indicating which edges carry flow, and that testing $x$ either returns a solution of cost at most $2\kappa B$ or separates $x$ from $\mathcal Q_B$. Applying Lemma~\ref{lem:constructive-feasibility} and doubling $B$ yields the claimed approximation.

\paragraph{Region processing.}
Given a rational vector $x\in[0,1]^E$ with $c(x)\le B$, we apply Lemma~\ref{lem:trees-small-boundaries} and process the resulting regions $V_i$ from the leaves to the root. If $V_i$ has insufficient remaining facility capacity, stop. Otherwise, we apply $\varUpsilon$ to the local instance on $G[V_i]$ with tree $\T_i$ whenever $V_i$ has positive remaining demand.

For each local solution, we find an integral cycle-free flow $g_i$ on $G[V_i]$, discard facilities with zero receipt, and let $A_i$ be the resulting cable cost plus the opening costs paid in that call. Set $g_i=0$ and $A_i=0$ when no demand remains. After each call, we subtract the served demand and facility receipts, and set the opening costs paid in that call to zero. A run is completed if every region is processed.

\paragraph{The inequalities.}
To find suitable edge values, we use Lemma~\ref{lem:constructive-feasibility}. For each candidate $x$, we either construct a low-cost solution by processing the regions or return a violated inequality satisfied by the vector indicating which edges carry flow in every feasible integral solution of cost at most $B$. We first define these inequalities. Let $\mathcal Q_B$ be the set of edge vectors $z$ satisfying
\begin{align}
&0\le z_e\le1\qquad(e\in E),\label{eq:global-bounds}\\*
 &\sum_e c_ez_e\le B,\label{eq:global-budget}\\*
 &z(\partial W)\ge1\qquad\bigl(W\subseteq V:\ \sum_{v\in W}d_v>\sum_{r\in U\cap W}b_r\bigr),\label{eq:global-capacity}\\*
 &\sum_i A_i\le\alpha B+\beta\sum_i c(\T_i)
   +(1+\alpha)\sum_i A_i z(\partial V_i)
   \qquad(\text{for every completed run}).\label{eq:global-cost}
\end{align}
Constraint~\eqref{eq:global-cost} applies to every completed run of the region processing above. For each such run with edge values $x$, its regions $V_i$, trees $\T_i$, and costs $A_i$ are fixed; only $z$ is a variable.
The regions, trees, and processing orders have finitely many choices. Each integral cycle-free flow $g_i$ carries at most $d_J$ units on any edge. Thus each $A_i$ is a sum of input edge and opening costs with nonnegative integer coefficients at most $d_J$, so it has finitely many possible values with polynomial encoding length. The inequalities therefore form a finite family with a known polynomial bound on their encoding lengths in $\enc$ and the encoding length of $B$. Hence $\mathcal Q_B$ satisfies the size requirements of Lemma~\ref{lem:constructive-feasibility}.

\paragraph{Validity of the inequalities.}
We show that every feasible integral solution of cost at most $B$ has a vector $p$ in $\mathcal Q_B$. Fix such a solution with flow $f_B$, and let $p_e=1$ if $f_{B,e}\ne0$ and $p_e=0$ otherwise. Then $p\in[0,1]^E$ and $\sum_e c_ep_e\le C_Q(f_B)\le B$. If $W$ has more demand than facility capacity, as in \eqref{eq:global-capacity}, some demand must cross its boundary, so $p(\partial W)\ge1$. Thus $p$ satisfies \eqref{eq:global-bounds}--\eqref{eq:global-capacity}. We next consider \eqref{eq:global-cost}.

Consider any completed run. We maintain a feasible flow $f$ for the remaining demand (initially $f_B$) and compare its cost before and after processing each region. Before processing $V_i$, let $b'_r$ be the remaining capacity of facility $r$. The local flow $g_i$ serves part of the demand served by $f$ and is zero outside $V_i$.

Apply Lemma~\ref{lem:residual-flow} to $f$ and $g_i$, and cancel cycles in the resulting flow $h$. For each facility $r$, the lemma gives $q_r^h\le(q_r^f-q_r^{g_i})_+\le\min\{q_r^f,b'_r-q_r^{g_i}\}$. Thus $h$ serves the remaining demand, uses only facilities used by $f$, and respects the remaining capacities. We replace $f$ by $h$ after processing $V_i$.

Before processing $V_i$, let $\Psi$ be the cable cost of $f$ plus the current opening costs of its receiving facilities. Let $\Delta\Psi_i$ be the change in this cost after replacing $f$ by $h$ and updating the opening costs. Opening costs do not increase, and $C_Q(h)\le C_Q(f)+C_Q(g_i)$ by \eqref{eq:subadd} and Lemma~\ref{lem:residual-flow}. Hence $\Delta\Psi_i\le C_Q(h)-C_Q(f)\le C_Q(g_i)\le A_i$, since $A_i$ includes the cable cost $C_Q(g_i)$.

We consider two cases. Suppose $p(\partial V_i)\ge1$. The bound $\Delta\Psi_i\le A_i$ gives $A_i+\alpha\Delta\Psi_i\le(1+\alpha)A_i$.

Now suppose $p(\partial V_i)=0$. Every earlier region is contained in $V_i$ or disjoint from it, so its local flow has no edge across $\partial V_i$. By $|h_e|\le|f_e-g_{i,e}|$ in Lemma~\ref{lem:residual-flow}, each update from $f$ to $h$ therefore preserves the absence of flow across this boundary. Since $f_B$ has no flow across the boundary, neither does $f$. Its restriction to $V_i$ is feasible for the local instance. Let $\Phi_i$ and $C_i$ be its opening and cable costs.

After $V_i$ is processed, no demand remains there. $h$ has no flow across $\partial V_i$, so flow conservation implies that its facilities in $V_i$ receive no demand. Since $h$ has no cycles, it also has no flow on edges inside $V_i$. Thus the costs $\Phi_i+C_i$ disappear from $\Psi$. Outside $V_i$, the local flow $g_i$ is zero, so $|h_e|\le|f_e|$, and facility receipts and opening costs do not increase. Hence $\Delta\Psi_i\le-(\Phi_i+C_i)$. By the assumption on $\varUpsilon$ in Lemma~\ref{lem:global}, $A_i\le\alpha(\Phi_i+C_i)+\beta c(\T_i)$. Thus $A_i+\alpha\Delta\Psi_i\le\beta c(\T_i)$.

Together, the two cases give
\[
 A_i+\alpha\Delta\Psi_i\le\beta c(\T_i)+(1+\alpha)A_i p(\partial V_i).
\]

Let $\Psi_{\mathrm{initial}}$ and $\Psi_{\mathrm{final}}$ be the values before and after processing all regions. Summing over the regions gives $\sum_i A_i+\alpha(\Psi_{\mathrm{final}}-\Psi_{\mathrm{initial}})\le\beta\sum_i c(\T_i)+(1+\alpha)\sum_i A_i p(\partial V_i)$. Since $\Psi_{\mathrm{initial}}\le B$ and $\Psi_{\mathrm{final}}\ge0$, \eqref{eq:global-cost} holds for $z=p$. Therefore $p\in\mathcal Q_B$. In particular, $\mathcal Q_B$ is nonempty whenever $B\ge\OPT$.

\paragraph{Testing a candidate.}
We now verify the oracle requirements of Lemma~\ref{lem:constructive-feasibility}. Given a rational vector $x$, return a violated bound in \eqref{eq:global-bounds} if $x\notin[0,1]^E$, or \eqref{eq:global-budget} if $c(x)>B$. Otherwise, run the region processing described above. Before processing $V_i$, let $d'_v$ and $b'_r$ be the remaining demand of client $v$ and capacity of facility $r$, respectively. If $\sum_{v\in V_i}d'_v>\sum_{r\in U\cap V_i}b'_r$, every earlier region is contained in $V_i$ or disjoint from it. Each earlier call inside $V_i$ removed the same amount of demand and facility capacity from $V_i$, so the original demand in $V_i$ is greater than its original facility capacity. Return \eqref{eq:global-capacity} for $W=V_i$. Since the input instance is feasible, $V_i$ is not the root, and $x(\partial V_i)<\delta<1$. Thus this inequality is violated.

If all regions are processed, set $A=\sum_iA_i$. The capacity and opening-cost updates ensure that their solutions together are feasible, pay each opening cost once, and cost at most $A$. If $A\le2\kappa B$, return the solution. Otherwise, return \eqref{eq:global-cost} for this run. Specifically, Lemma~\ref{lem:trees-small-boundaries} and $c(x)\le B$ give $\sum_i c(\T_i)\le4(1+\alpha)B$. Thus $\alpha B+\beta\sum_i c(\T_i)\le\kappa B<\frac{A}{2}$. The root boundary is empty and every other boundary has $x$-value less than $\delta$, so $(1+\alpha)\sum_iA_i x(\partial V_i)\le(1+\alpha)\delta A=\frac{A}{2}$. The inequality is therefore violated by $x$. When a solution is returned, we do not need to check whether $x$ satisfies all constraints in \eqref{eq:global-bounds}--\eqref{eq:global-cost}.

\paragraph{Searching over the budgets.}
Start with the least positive edge or opening cost, which is at most $\OPT$. For each budget $B$, apply Lemma~\ref{lem:constructive-feasibility} using the tests above as the oracle; if $\mathcal Q_B$ is empty, double $B$ and repeat. Since $\mathcal Q_B$ is nonempty when $B\ge\OPT$, the search returns a solution of cost at most $2\kappa B<4\kappa\OPT$. Each test makes at most $\size{V}$ calls to $\varUpsilon$, and an all-open direct assignment has cost at most $\sum_{r\in U}\phi_r+d_J\max_{e\in E}c_e$, which has polynomial encoding length. Thus the search uses polynomially many budgets and runs in time $\poly(\enc)$.
\end{proof}

It remains to construct a local algorithm $\varUpsilon$ with constant $\alpha$ and $\beta$.

\section{Local Cable Construction}\label{sec:local}
In this section, we call the clients sources.
To apply the global-to-local reduction, we first simplify facilities with small capacities. If $b_r\le Q$, all demand assigned to facility $r$ fits on a single cable. We can therefore replace facility $r$ by free capacity at a new leaf and represent its opening cost by the cost of the pendant edge.

\begin{lemma}\label{lem:leaf}
For any facility $r\in U$ with $b_r\le Q$, replace $r$ by a free facility of capacity $b_r$ at a new leaf $r'$ joined to $r$ by an edge of cost $\phi_r$. In the resulting shortest-path metric, solutions convert in either direction without increasing cost.
\end{lemma}
\begin{proof}
Given a solution of the original instance, send any positive receipt of facility $r$ along its new pendant edge to facility $r'$ on one cable. Conversely, given a solution of the modified instance, expand its edges into shortest paths in the modified graph and cancel flow in opposite directions; \eqref{eq:subadd} bounds the resulting cable cost by that of the metric solution. Since the leaf has no demand, its pendant flow equals its receipt and is at most $Q$. Discard any extra copies of the pendant cable. If the leaf has positive receipt, move this receipt back to facility $r$ and replace the single pendant cable by the opening payment $\phi_r$. Adding leaves does not change the shortest-path costs between original nodes, so the cost does not increase.
\end{proof}

Applying Lemma~\ref{lem:leaf} once to the original instance ensures that every nonfree facility has capacity greater than $Q$. In every subsequent local subproblem, each unopened nonfree facility retains its capacity, while each previously opened facility has zero opening cost and thus contributes free capacity. When $d_J>0$, we may cap the cable capacity and all facility capacities at $d_J$, as in Section~\ref{sec:tools}, and denote the resulting cable capacity again by $Q$. Each nonfree facility then has capacity at least $Q$.
If $d_J=0$, we simply return the empty solution.

To establish the local guarantee required by Lemma~\ref{lem:global}, it therefore suffices to consider the following restriction of the local subproblem.

\begin{definition}[Local problem]\label{def:local}
The input is a feasible residual H-CCFL instance with arbitrary free capacities and nonfree facilities of capacity at least $Q$, together with a tree $\T$ connecting its sources. The objective is to minimize the total opening and cable cost.
\end{definition}

Each node has at most one facility, which is either free or nonfree.
We may assume $\OPT>0$. Set $\tau=c(\T)$. For each node $i$, let $\hat b_i$ be its free capacity (zero if none). We also let $b_i$ denote the capacity of each nonfree facility $i$.
The following lemma converts a feasible fractional flow into a cable solution when a tree connects the sources and a set of facilities.

\begin{lemma}\label{lem:compatible}
Let $f$ be a feasible rational flow from sources with integer demands to a given set of facilities with integer capacities, and let $H$ be a tree connecting all sources and these facilities. There exists a polynomial-time algorithm that constructs a feasible integral cable solution with cable cost at most $\frac{L(f)}{Q}+c(H)$.
\end{lemma}
\begin{proof}
The proof is given in Section~\ref{app:compatible}.
\end{proof}

\paragraph{Overview of the local algorithm.}
The input tree $\T$ connects the sources. By Lemma~\ref{lem:compatible}, it suffices to find a tree $H$ connecting the sources and a set of facilities, and a feasible flow $f$ to these facilities, while bounding $c(H)$ and $\frac{L(f)}{Q}$. The flow need not lie on $H$.

We first introduce a fractional relaxation under a budget, use Lemma~\ref{lem:trees-small-boundaries} to connect some free facilities to the sources, and apply a quota-tree algorithm to group the remaining free facilities of capacity less than $Q$ (Section~\ref{sec:ingredients}). Each group serves as a facility in an auxiliary CFL instance. We then use a maximum flow to assign part of the demand to free facilities and nonfree facilities with sufficiently large opening fractions, and for the remaining demand either return a violated inequality or obtain a rounded assignment to selected group and nonfree facilities (Section~\ref{sec:residual}).

Next, we combine this assignment with the partial assignment, merge small receipts to connect the selected facilities at bounded cost, and use a quota tree to supply capacity for the small amount of demand left unassigned; a minimum-cost flow and Lemma~\ref{lem:compatible} then yield the cable solution (Section~\ref{sec:rounding}). Finally, we bound the total cost and search over budgets to establish the local guarantee (Section~\ref{sec:local-cost}).

\subsection{Fractional Relaxation and Grouping Free Facilities}\label{sec:ingredients}
For each budget $B$, we first introduce a fractional relaxation. We then use its edge values to connect some free facilities to the sources and group the small free facilities outside the resulting forest. Each group can be treated as a facility; less than $Q$ units of small free capacity may remain ungrouped.

\paragraph{Fractional relaxation.}
For the relaxation, an entry $i$ is either a free facility of capacity $\bar b_i=\hat b_i$ or a nonfree facility of capacity $\bar b_i=b_i$. We omit entries of zero capacity. Let $c(i,J)=\min_{s\in J}c(i,s)$.

Let $X_{i,s}$ denote the fractional demand assigned from source $s$ to entry $i$, $Y_i$ the opening fraction of a nonfree facility, and $z_e$ the fractional value of edge $e$. Set $\hat c=\frac{c}{Q}$, $\ell=\ell(X)=\sum_{i,s}\hat c(i,s)X_{i,s}$, $\Phi'=\sum_{i\text{ nonfree}}\phi_iY_i$, and $\Omega=\sum_ec_ez_e$. These are the scaled assignment cost, fractional opening cost, and fractional edge cost, respectively. We consider the following constraints:
\begin{align}
 \text{find}\quad & (X,Y,z) \notag\\*
 \text{subject to}\quad &\sum_iX_{i,s}=d_s\quad(s\in J),\qquad X_{i,s}\ge0,\label{eq:aggregate}\\*
 &\sum_sX_{i,s}\le \bar b_i\quad(i\text{ free}),\label{eq:free-capacity}\\*
 &\sum_sX_{i,s}\le b_iY_i,\quad X_{i,s}\le d_sY_i
       \quad(i\text{ nonfree},\ s\in J),\label{eq:nonfree-capacity}\\*
 &0\le Y_i\le1\quad(i\text{ nonfree}),\qquad
  0\le z_e\le1\quad(e\in E),\label{eq:free-edge-bounds}\\*
 &\Omega+\Phi'+\ell\le B.\label{eq:witnessbudget}
\end{align}
The assignment constraints serve all demand within the entry capacities, and \eqref{eq:witnessbudget} bounds the total cost by $B$. We let $\mathcal P_B$ denote the convex hull of the integer points satisfying these constraints for which each free entry of capacity less than $Q$ with positive receipt is connected to a source by edges of positive $z$-value.

Consider any feasible local solution with opening cost $\Phi$ and cable cost $C$. Choose an integral cycle-free flow for this solution and replace its source--facility paths by direct assignments to obtain $X$. By the triangle inequality and \eqref{eq:subadd}, $\ell\le C$. Let $Y$ be the opening indicators of the nonfree facilities and $z$ the incidence vector of the flow support, so $\Omega\le C$. The opening indicators satisfy $\Phi'\le\Phi$. Every entry with positive receipt is connected to a source by the support; if the entry is at that source, the connection uses no edges. Thus the resulting point belongs to $\mathcal P_B$ when $B\ge\Phi+2C$, so $\mathcal P_B$ is nonempty.

Since $0\le X_{i,s}\le d_s$ and $Y_i,z_e\in\{0,1\}$ at the integer points defining $\mathcal P_B$, their convex hull has a polynomial bound on its facet encoding lengths. Thus $\mathcal P_B$ satisfies the size requirements of Lemma~\ref{lem:constructive-feasibility}.
Given a rational candidate $(X,Y,z)$, we first test \eqref{eq:aggregate}--\eqref{eq:witnessbudget}. If a constraint is violated, we return the corresponding inequality, or the violated side of an equality. Each such inequality is valid for $\mathcal P_B$. Otherwise, we use $(X,Y,z)$ to construct the forest and groups below.

\paragraph{Connecting entries to sources.}
We first use the candidate edge values $z$ to connect some free entries to the sources.
Add a root $r_0$ joined to every source by an edge of cost zero and $z$-value one. Apply Lemma~\ref{lem:trees-small-boundaries} to this graph and edge vector $z$ with a rational parameter $\delta\in(0,1)$ to be chosen later. Any two different node sets of the partition are separated by the boundary of a nonroot region, whose $z$-value is less than $\delta$. Since $\delta<1$ and each edge from $r_0$ to a source has $z$-value one, all sources belong to the node set containing $r_0$.

Take the tree spanning this node set and delete $r_0$ and its incident edges, retaining all other nodes. The resulting forest $F$ contains all sources. Each component contains a source because every neighbor of $r_0$ is a source. The added edges have zero cost, so $c(F)\le\frac{2\Omega}{\delta}$ by Lemma~\ref{lem:trees-small-boundaries}.

We next consider the free entries of capacity less than $Q$ whose locations are outside $F$, and denote their set by $I_{\rm sm}$. For each $i\in I_{\rm sm}$, the node sets containing $i$ and $r_0$ are separated by the boundary of some nonroot region in the rooted tree from Lemma~\ref{lem:trees-small-boundaries}. Let $R_i$ be the side of this boundary containing $i$.
Since all sources belong to the same node set as $r_0$, $R_i$ contains no source or $r_0$, and $\partial R_i$ consists only of original edges.  Hence, by Lemma~\ref{lem:trees-small-boundaries},
\begin{equation}\label{eq:small-cut}
 z(\partial R_i)<\delta,\qquad i\in I_{\rm sm}.
\end{equation}
The sets $R_i$ remain fixed in the following comparison. At any integer point $(X^0,Y^0,z^0)$ defining $\mathcal P_B$, an entry $i\in I_{\rm sm}$ with $\sum_sX^0_{i,s}>0$ is connected to a source by edges with positive $z^0$-value. This connection crosses $\partial R_i$, so $z^0(\partial R_i)\ge1$. We use these sets and their boundaries below to bound the receipts of groups of entries.

\paragraph{Forming groups.}
In the rooted weighted quota-tree problem, each node has a nonnegative integer prize, and the objective is to find a minimum-cost tree containing the root and nodes whose total prize reaches a given quota. We use the $(2+\varepsilon)$-approximation algorithm of \citet{conf:friggstad:WADS:25} with $\varepsilon=1$ on edge-weighted graphs, returning a tree in the input graph. We apply it to the original graph, without the added root $r_0$.

Initially all entries in $I_{\rm sm}$ are ungrouped. While their total ungrouped capacity is at least $Q$, assign each node a prize equal to the capacity of its ungrouped entry in $I_{\rm sm}$, if any. For each source, apply the quota-tree algorithm with that source as root and quota $Q$. Choose one of minimum cost among the returned trees, and select whole ungrouped entries on it until their total capacity first reaches $Q$.
Each iteration groups at least one previously ungrouped entry. Thus there are at most $|I_{\rm sm}|$ iterations, each using at most $|J|$ calls to the quota-tree algorithm.

Let $\G$ be the selected group, $b_{\G}$ its total capacity, and $s_{\G}$ the root of the chosen tree. Let $\T_{\G}$ be the subtree of that tree connecting $s_{\G}$ and the locations of the entries in $\G$, and set $c_{\G}=c(\T_{\G})$. Its cost is at most that of the chosen tree. Since each entry has capacity less than $Q$, we have $Q\le b_{\G}<2Q$. Mark the entries of $\G$ as grouped and continue. Different trees may share nodes and edges, but the groups are disjoint sets of entries. When the procedure stops, the remaining ungrouped capacity is less than $Q$.

\begin{lemma}\label{lem:cheap-groups}
Let $S\subseteq I_{\rm sm}$ have total capacity at least $2Q$. If a tree of cost $t$ connects the entries of $S$ and a source, then the groups $\G$ with $c_{\G}\le3t$ contain entries of $S$ of total capacity at least $Q$.
\end{lemma}
\begin{proof}
Consider the first iteration producing a group $\G$ with $c_{\G}>3t$. At the start of that iteration, the ungrouped capacity of $S$ is less than $Q$: otherwise, the tree in the lemma is a feasible comparison for the quota-tree call rooted at its source, so the selected group costs at most $3t$. Thus more than $Q$ of capacity in $S$ belongs to earlier groups, each of cost at most $3t$.

If no such iteration occurs, the procedure stops with less than $Q$ total capacity ungrouped. Hence more than $Q$ of the capacity in $S$ belongs to groups, all of cost at most $3t$.
\end{proof}

\subsection{Residual Assignment}\label{sec:residual}
We divide the assignment into two parts. First, we assign part of the demand directly to entries that do not require grouping and can be opened at bounded cost. Specifically, let $I$ consist of all free entries outside $I_{\rm sm}$ (which either have capacity at least $Q$ or already lie on $F$) together with all nonfree facilities satisfying $Y_i\ge\frac{1}{4}$.

Second, for the remaining demand, we construct an auxiliary CFL instance $\I_{\mathrm{aux}}$ whose facilities represent the groups from Section~\ref{sec:ingredients} and the remaining nonfree facilities. Its demands are determined later.
To account for the connection cost of each group $\G$, introduce a new leaf joined to $s_{\G}$ by an edge of cost $c_{\G}$. We extend $c$ to these leaves using shortest-path costs in the resulting graph; the costs between original nodes remain unchanged. Place a group facility at this leaf with capacity $b_{\G}$, opening cost zero, and opening fraction $y_{\G}=1$, so assigning one unit from source $s$ to it has scaled cost $\hat c(s,\G)=\frac{c(s,s_{\G})+c_{\G}}{Q}$. For a group facility $j$, let $c(j,J)=\min_{s\in J}c(j,s)$, as for an original facility. Include also each nonfree facility $i\notin I$ at node $i$, with its original capacity $b_i$, opening cost $\phi_i$, and opening fraction $y_i=4Y_i<1$.

Each group or nonfree facility $j$ has capacity $b_j\ge Q$ (by construction for groups and by Definition~\ref{def:local} for nonfree facilities). In the CFL instance used for rounding, we use only $\frac{b_j}{4}$ of this capacity. Rounding may assign up to twice this amount, and we later double the assignment, so the final assignment respects $b_j$.

We set $\delta=\frac{1}{12}$ in the construction of $F$ and the sets $R_i$ in Section~\ref{sec:ingredients}; as shown in the proof of the following lemma, this choice ensures that the fractional demand routed to any group $\G$ stays within $\frac{b_{\G}}{4}$.

\begin{lemma}[Residual assignment]\label{lem:residual}
For a candidate $(X,Y,z)$ satisfying \eqref{eq:aggregate}--\eqref{eq:witnessbudget}, a polynomial-time algorithm either returns a polynomial-size inequality valid for $\mathcal P_B$ and violated by $(X,Y,z)$, or finds: (1) a nonnegative partial assignment $\widetilde X$ to $I$ respecting the capacities $\bar b_i$ with $\ell(\widetilde X)\le4\ell$, and (2) a set of selected group and nonfree facilities of $\I_{\mathrm{aux}}$ and a nonnegative assignment $x$ to them respecting their capacities $b_j$.
The demand of $\I_{\mathrm{aux}}$ at source $s$ is $D_s:=d_s-\sum_{i\in I}\widetilde X_{i,s}$. The assignment $x$ serves nonnegative demands $d'_s\le D_s$ with $\sum_{s\in J}(D_s-d'_s)\le8Q$. The opening cost $O'$ of these selected facilities and the scaled assignment cost $L'$ of $x$ satisfy
\begin{equation}\label{eq:aux-cost}
 O'+L'\le36\left(4\sum_{i\notin I,\,i\text{ nonfree}}\phi_iY_i+\Lambda\right),
\end{equation}
where $\Lambda=42\Omega+44\tau+8\ell+\ell(\widetilde X)\le42\Omega+44\tau+12\ell$.
\end{lemma}
\begin{proof}
The proof is given in Section~\ref{app:residual}.
\end{proof}

\subsection{Cable Construction}\label{sec:rounding}
Suppose Lemma~\ref{lem:residual} returns a partial assignment $\widetilde X$ to $I$ and an assignment $x$ to the selected facilities of $\I_{\mathrm{aux}}$. The groups are disjoint and contain no entry of $I$, and the nonfree facilities used by $x$ are outside $I$; thus these assignments respect the original capacities together. We proceed in three steps. We first connect the facilities used by $\widetilde X$ and $x$ to the sources. We then find enough capacity to serve the unassigned demand. Finally, we construct a tree and a flow and apply Lemma~\ref{lem:compatible} to obtain a cable solution.

\paragraph{Connecting the selected facilities.}
Let $\widetilde X_F$ be the part of $\widetilde X$ assigned to entries whose locations lie on $F$, and let $\widetilde X_{\rm out}$ be the remaining part. The forest $F$ already connects the entries used by $\widetilde X_F$ to the sources, so we keep this part unchanged. We combine $\widetilde X_{\rm out}$ with the assignment $x$ and connect the facilities they use.
The assignment $\widetilde X_{\rm out}$ can use free entries and nonfree facilities in $I$ outside $F$, while $x$ can use group facilities and nonfree facilities outside $I$. Each of these facilities has capacity at least $Q$: this holds for free entries in $I$ outside $F$ by the definition of $I$, for group facilities by their construction, and for nonfree facilities by Definition~\ref{def:local}.

A facility receiving much less than $Q$ may be expensive to connect even when its scaled assignment cost is small. We therefore merge some of these receipts before connecting the facilities.

\begin{lemma}\label{lem:large}
Suppose a feasible assignment sends demand to a set of facilities of capacity at least $Q$, at scaled assignment cost $L$. There exists a polynomial-time algorithm that retains a subset of these facilities and serves all but less than $\frac{Q}{2}$ of this demand, with scaled assignment cost at most $3L+2\tau$. The retained facilities can be connected to nearest sources at total cost at most $2L$, and their opening cost does not increase.
\end{lemma}
\begin{proof}
We first keep facilities receiving at least $\frac{Q}{2}$ units and group the smaller positive receipts along $\T$.
Let $q_j$ be the receipt of a facility considered here, and choose a nearest source $s_j$ for it, so $c(j,J)=c(j,s_j)$ by the definition in Section~\ref{sec:residual}. We keep every facility with $q_j\ge\frac{Q}{2}$ and its assignment. To combine the smaller receipts, we traverse $\T$ using each edge twice and order the facilities by the first visits to their sources $s_j$, keeping facilities with the same nearest source consecutive. We group whole receipts in this order until each group first reaches $\frac{Q}{2}$. Since each receipt is less than $\frac{Q}{2}$, every complete group has total receipt $q\in[\frac{Q}{2},Q)$. We omit the final incomplete group, whose total receipt is less than $\frac{Q}{2}$.

We next assign the demand of each complete group to one facility and bound the additional cost of connecting facilities to sources.
For each complete group, retain a facility $h$ minimizing $c(j,J)$ among the facilities in the group and assign the group's $q$ units of demand to $h$. This assignment respects the capacity of $h$ because $q<Q$ and its capacity is at least $Q$. To bound the new scaled assignment cost, follow each original assignment to its facility $j$, then go to $s_j$, along $\T$ to $s_h$, and finally to $h$. The two portions connecting facilities to their nearest sources have scaled cost at most $\frac{1}{Q}\sum_jq_j[c(j,J)+c(h,J)]\le\frac{2}{Q}\sum_jq_jc(j,J)$ by the choice of $h$, where the sums are over the group.

We then bound the cost along $\T$ and the total scaled assignment cost.
For each group, the part of the traversal from its first to its last source connects all its sources. These parts do not overlap in the traversal, whose total cost is $2\tau$. Each group has less than $Q$ units, so its scaled assignment cost along this part is at most its cost. The total scaled cost along $\T$ is therefore at most $2\tau$. Also, since each unit originally assigned to $j$ has connection cost at least $c(j,J)$, we have $\frac{1}{Q}\sum_jq_jc(j,J)\le L$, where the sum is over all facilities considered in the lemma. The original assignment and the two added portions therefore have total scaled cost at most $3L+2\tau$.

Finally, we bound the cost of connecting the retained facilities.
For a facility kept with its original assignment, $c(j,J)\le\frac{2q_jc(j,J)}{Q}$. For a complete group, the choice of $h$ gives $c(h,J)\le\frac{\sum_jq_jc(j,J)}{q}\le\frac{2}{Q}\sum_jq_jc(j,J)$. Summing these bounds over all retained facilities, their total connection cost to nearest sources is at most $2L$.
Note that only a subset of the original facilities is retained, so the opening cost does not increase.
\end{proof}

We apply Lemma~\ref{lem:large} to the assignment consisting of $\widetilde X_{\rm out}$ and $x$, whose scaled cost is $L=L'+\ell(\widetilde X_{\rm out})$.

Recall that $x$ can assign demand to a group facility at the leaf for group $\G$. We then replace each retained group facility by its entries. The group leaf is joined only to source $s_{\G}$ by an edge of cost $c_{\G}=c(\T_{\G})$, by the definition of $c_{\G}$ in Section~\ref{sec:ingredients}. Thus its connection cost to a nearest source is $c_{\G}$. In the original graph, we use $\T_{\G}$ in place of this auxiliary edge and distribute the leaf's receipt among the entries of $\G$ within their capacities.
For a source $s$, the cost of connecting $s$ to the group leaf is $c(s,s_{\G})+c_{\G}$. This also bounds the cost of a path from $s$ through $s_{\G}$ to any entry of $\G$ on $\T_{\G}$. By the triangle inequality, the direct connection from $s$ to the entry costs no more. The replacement therefore does not increase the scaled assignment cost. Moreover, $\T_{\G}$ connects all entries of $\G$ to $s_{\G}$ at the same cost $c_{\G}$ as the auxiliary edge. Since the groups and the other retained facilities represent disjoint entries, all original capacities remain respected.

By Lemma~\ref{lem:large}, the part of the assignment obtained from $\widetilde X_{\rm out}$ and $x$ has scaled cost at most $3L+2\tau$, and its entries outside $F$ can be connected to sources at cost at most $2L$.
Every nonfree facility in $I$ has $Y_i\ge\frac{1}{4}$, and the opening costs of the facilities selected by $x$ outside $I$ are included in $O'$. The following lemma summarizes the costs of the resulting assignment and connections.

\begin{lemma}\label{lem:partial-assignment}
The resulting partial assignment has scaled assignment cost at most $\ell(\widetilde X_F)+3L+2\tau$, and the opening cost of its nonfree facilities is at most $4\sum_{i\in I,\,i\text{ nonfree}}\phi_iY_i+O'$. The entries used by the assignment and located outside $F$ can be connected to sources at total cost at most $2L$.
\end{lemma}

\paragraph{Serving the remaining demand.}
Let $\Delta$ be the total demand left unassigned by the preceding construction. At most $8Q$ units were left unassigned by Lemma~\ref{lem:residual}, and less than $\frac{Q}{2}$ are omitted by Lemma~\ref{lem:large}, so $\Delta<9Q$. We now serve this demand.
If $\Delta>0$, we use the $3$-approximate quota-tree algorithm~\citep{conf:friggstad:WADS:25} to find enough capacity to serve all $\Delta$ units of unassigned demand.
We seek total capacity at least $d_J$, so that at least $\Delta$ remains after the preceding assignment uses $d_J-\Delta$ units.
Specifically, add a root joined to all sources by zero-cost edges, and give each original node a prize equal to its free capacity. For each nonfree facility $i$, add a leaf at its location with edge cost $\phi_i$ and prize $b_i$.
We then apply the quota-tree algorithm with quota $d_J$.

To bound the returned tree's cost, consider any feasible local solution with opening cost $\Phi$ and cable cost $C$. The support of a cycle-free flow for this solution, together with the root edges and the leaves of receiving nonfree facilities, forms a connected subgraph containing the root and prize at least $d_J$. Its cost is at most $\Phi+C$, so the quota-tree algorithm returns a tree of cost at most $3(\Phi+C)$.

We then delete the root and the added facility leaves, adding missing sources as isolated nodes.
Let $H_0$ be the resulting forest and $P_0$ the facilities whose leaves were in the tree. Since the root is joined only to sources, each component of $H_0$ contains a source. The location of every facility in $P_0$ lies in $H_0$ because its leaf was attached to that location. The free capacity at nodes of $H_0$ and the capacity of $P_0$ together are at least $d_J$, and
\begin{equation}\label{eq:initial-tree-cost}
 \Phi(P_0)+c(H_0)\le3(\Phi+C).
\end{equation}

If $\Delta=0$, set $H_0$ and $P_0$ to be empty.
If $\Delta>0$, the assignment so far uses $d_J-\Delta$ units of capacity, while $H_0$ and $P_0$ provide at least $d_J$ units. Thus at least $\Delta$ units of the latter capacity remain unused. We assign the unassigned demand to this unused capacity.
Each component of $H_0$ contains a source, and $\T$ connects all sources, so $\T\cup H_0$ is connected and reaches every facility in $P_0$. Route this demand on a spanning tree of this union. Each edge carries at most $\Delta<9Q$ units, so the scaled cost of this flow is at most $9(\tau+c(H_0))$.

\paragraph{Constructing the tree and cable flow.}
We now construct a tree $H$ in the original graph. The retained connections to sources constructed above include $\T_{\G}$ for each retained group. Take a spanning tree of their union with $\T$, $F$, and $H_0$. This union is connected because each of its parts meets a source and $\T$ connects all sources. Thus $H$ connects all sources, the retained entries, the free entries at nodes of $H_0$, and the facilities in $P_0$. We open the nonfree facilities used by the partial assignment and the facilities in $P_0$. We then find a minimum-cost assignment of all demand to these facilities, the retained free entries, and the free entries at nodes of $H_0$, counting each entry only once.

The assignments constructed above show that this problem is feasible. Replacing their paths by direct connections cannot increase their cost, by the triangle inequality. Since demands and capacities are integers, the minimum-cost assignment can be chosen integral. We send its demand along shortest paths in the original graph and cancel opposing flows to obtain a flow $f$. Its cost $\frac{L(f)}{Q}$ is at most the scaled assignment cost. Thus, by Lemma~\ref{lem:partial-assignment} and the bound for serving the unassigned demand,
\begin{equation}\label{eq:local-flow-cost}
 \frac{L(f)}{Q}\le\ell(\widetilde X_F)+3(L'+\ell(\widetilde X_{\rm out}))+2\tau+9(\tau+c(H_0)).
\end{equation}
Finally, we apply Lemma~\ref{lem:compatible} to $f$ and $H$ to obtain the cable solution.

\subsection{Cost Analysis and Proof of the Main Theorem}\label{sec:local-cost}
We now establish the guarantee of the local algorithm.

\begin{lemma}[Local approximation]\label{lem:local}
Given $\T$, the local algorithm computes in $\poly(\enc)$ time a feasible solution of cost at most $15198\Phi+30366C+7932c(\T)$ for every feasible local solution with opening cost $\Phi$ and cable cost $C$.
\end{lemma}
\begin{proof}
Fix a feasible local solution with opening cost $\Phi$ and cable cost $C$.

\paragraph{Cost bound for a fixed budget.}
Let $\operatorname{sol}$ be the cost of the solution constructed in Section~\ref{sec:rounding} for a candidate $(X,Y,z)$ passing the tests in Sections~\ref{sec:ingredients}--\ref{sec:residual}. By Lemma~\ref{lem:partial-assignment}, the total opening cost is at most $4\sum_{i\in I,\,i\text{ nonfree}}\phi_iY_i+O'+\Phi(P_0)$, since any facility used by the partial assignment and included in $P_0$ is opened only once. The flow $f$ satisfies \eqref{eq:local-flow-cost}. The tree $H$ has cost at most $\tau+c(F)+2(L'+\ell(\widetilde X_{\rm out}))+c(H_0)$ by definition.
We combine these bounds using Lemma~\ref{lem:compatible}.

By Lemma~\ref{lem:residual}, $\ell(\widetilde X)\le4\ell$, so $\ell(\widetilde X_F)+5\ell(\widetilde X_{\rm out})\le5\ell(\widetilde X)\le20\ell$. By Lemma~\ref{lem:trees-small-boundaries} and $\delta=\frac{1}{12}$, we have $c(F)\le24\Omega$. Finally, \eqref{eq:initial-tree-cost} gives $\Phi(P_0)+10c(H_0)\le10[\Phi(P_0)+c(H_0)]\le30(\Phi+C)$. Thus
\begin{equation}\label{eq:ledger}
\begin{aligned}
 \operatorname{sol}
 &\le4\sum_{i\in I,\,i\text{ nonfree}}\phi_iY_i+O'+5L'
      +20\ell+24\Omega+12\tau+30(\Phi+C)\\
 &\le720\Phi'+180\Lambda+20\ell+24\Omega+12\tau+30(\Phi+C)\\
 &\le720\Phi'+7584\Omega+2180\ell+7932\tau+30(\Phi+C)\\
 &\le7584B+7932\tau+30(\Phi+C),
\end{aligned}
\end{equation}
where the second inequality follows from $O'+5L'\le5(O'+L')$, \eqref{eq:aux-cost}, and $\sum_{i\text{ nonfree}}\phi_iY_i=\Phi'$, the third from $\Lambda\le42\Omega+44\tau+12\ell$ by Lemma~\ref{lem:residual}, and the last from \eqref{eq:witnessbudget}.

\paragraph{Searching over the budgets.}
Since $\OPT>0$, we initialize $B$ to the smallest positive edge or opening cost. For each budget $B$, we run the algorithm of Lemma~\ref{lem:constructive-feasibility} on $\mathcal P_B$. At each candidate $(X,Y,z)$, we test \eqref{eq:aggregate}--\eqref{eq:witnessbudget} and, if they hold, construct the forest and groups and apply Lemma~\ref{lem:residual}. If both tests pass, Sections~\ref{sec:ingredients}--\ref{sec:rounding} produce a solution satisfying \eqref{eq:ledger} and we stop; otherwise, we obtain a polynomial-size inequality valid for $\mathcal P_B$ and violated by $(X,Y,z)$. These steps provide the oracle for Lemma~\ref{lem:constructive-feasibility}. As in the proof of Lemma~\ref{lem:global}, we double $B$ if $\mathcal P_B$ is empty. By Section~\ref{sec:ingredients}, $\mathcal P_B$ is nonempty whenever $B\ge\Phi+2C$, so the search terminates by the first budget reaching $\Phi+2C$. As in the proof of Lemma~\ref{lem:global}, the number of budgets is polynomial; Lemma~\ref{lem:constructive-feasibility} then gives polynomially many calls per budget.

\paragraph{Approximation guarantee and running time.}
Since $\OPT>0$, the initial budget is at most $\Phi+C$. The search therefore returns a solution with $B<2(\Phi+2C)$. Substituting this bound and $\tau=c(\T)$ into \eqref{eq:ledger} gives the claimed cost. Feasibility follows from Sections~\ref{sec:ingredients}--\ref{sec:rounding}. By the budget analysis above and Lemma~\ref{lem:constructive-feasibility}, the search makes polynomially many calls; each call and the final construction run in $\poly(\enc)$ time, so the total running time is $\poly(\enc)$.
\end{proof}

We are now ready to prove the main result.

\begin{theorem}\label{thm:binary}
The H-CCFL admits a polynomial-time $\Gamma$-approximation for $\Gamma=2^{33}$, with arbitrary integer demands and facility capacities.
\end{theorem}
\begin{proof}
After the reduction in Lemma~\ref{lem:leaf}, Lemma~\ref{lem:local} gives a local solution of cost at most $2^{15}(\Phi+C)+2^{13}c(\T)$. Thus Lemma~\ref{lem:global} applies with $\alpha=2^{15}$ and $\beta=2^{13}$, giving approximation $4[\alpha+4\beta(1+\alpha)]=2^{32}+2^{18}<2^{33}=\Gamma$. The running time is $\poly(\enc)$ by Lemmas~\ref{lem:global}, \ref{lem:leaf}, and \ref{lem:local}.
\end{proof}

\section{Applications}\label{sec:applications}
We use Theorem~\ref{thm:binary} to obtain constant-factor approximations for the ARP, the MDSDVRP, and the H-CLR with depot capacities at least the vehicle capacity.
We use the graph $G$, clients $J$, and facilities $U$ from Section~\ref{sec:tools}, and may call the facilities in $U$ depots.
We first show how to construct trees from a cable solution. To represent these trees in polynomial size, we use tree records specifying the root facility, the tree's edges, the amounts served at its clients, and a binary repetition count.

\begin{lemma}\label{lem:fleetdecode}
Let $f$ be an integral cable flow from clients $J$ to depots $U$ with cable capacity $Q$, and let $q_r$ be the receipt at depot $r$. Suppose each depot $r$ with $q_r>0$ has an integer capacity $b_r\ge Q$ with $q_r\le b_r$. A polynomial-time algorithm serves all demand with trees of total cost at most $10C_Q(f)$, stored in $\O(nm)$ records. At most $\ceil{\frac{b_r}{Q}}$ trees are rooted at each receiving depot $r$, and none at other depots. Each tree serves at most $Q$ units in integer amounts at its clients and contains only its depot and the nodes it serves; the trees at $r$ together serve at most $b_r$ units. Moreover, if $b_r$ is divisible by $Q$ for every receiving depot $r$, the cost bound improves to $3C_Q(f)$.
\end{lemma}
\begin{proof}
When each $b_r$ is divisible by $Q$, the path construction and assignment method of \citet[Lemma~5]{preprint:zhao:arxiv:26a} can be used to obtain the $3C_Q(f)$ bound. We extend their method to general $b_r\ge Q$ by combining units with different depot labels and merging trees when needed. The proof is given in Section~\ref{app:fleetdecode}.
\end{proof}

\subsection{Airport and Railway Problem}\label{sec:ar}
In the ARP, let $G=(V,E)$ be a complete graph whose nodes represent the cities. The edge costs $c:E\to\mathbb{Q}_{\ge0}$ satisfy the triangle inequality. An airport may open at each $v\in V$ at cost $\phi_v\in\mathbb{Q}_{\ge0}$. Given airport capacity $K\in\mathbb{N}^+$, a solution partitions the cities into trees, each with exactly one open airport and at most $K$ cities, including its airport. The objective is to minimize the total tree and opening cost.

\citet{conf:adamaszek:STACS:16} introduced the ARP and developed a PTAS for the Euclidean plane with uniform opening costs. With arbitrary opening costs on general metrics, \citet{conf:adamaszek:STACS:18} obtained a $(\frac{4}{3}(2+\frac{1}{p}),1+p)$-approximation for any $0<p\le1$ with $pK\in\mathbb{Z}$: the cost is at most the first factor times the optimum under capacity $K$, and each tree contains at most $(1+p)K$ cities. For the Euclidean variant with Steiner nodes, they established a polynomial-time $(2+\frac{K}{K-1}+\varepsilon)$-approximation and a quasi-polynomial-time $(1+\varepsilon,1+\mu)$-approximation for any fixed $\varepsilon,\mu>0$. On general metrics, \citet{journal:salavatipour:JOCO:25} gave a $2$-approximation for uniform opening costs and an $\O(\log\size{V})$-approximation for arbitrary opening costs via randomized tree embeddings~\citep{journal:fakcharoenphol:JCSS:04}. They also studied structured metrics, such as those induced by planar graphs and graphs of bounded treewidth.

\citet{journal:jowhari:IPL:25} later studied the extension with unsplit integer demands, in which each city's demand must be served by one airport. They showed that it is NP-hard to approximate within any factor while respecting the original capacity, even when each demand is at most $\poly(\size{V})$. For polynomially bounded total demand, they showed that any polynomial-time $(\alpha,\beta)$-approximation for the original problem can be converted into a polynomial-time $(2\alpha,\beta+2)$-approximation for unsplit demands.

We next show that the ARP admits a constant-factor approximation.

\begin{theorem}\label{thm:ar}
The ARP admits a polynomial-time $3\Gamma$-approximation.
\end{theorem}
\begin{proof}
We may assume that $K\ge2$ since if $K=1$ every city must open an airport.
Given an instance $\I_{\mathrm{ARP}}$ of the ARP, we construct an instance $\I_{\mathrm{HCCFL}}$ of the H-CCFL in which every city supplies one unit and has a facility of capacity $K$ with its original opening cost, and the cable capacity is $K-1$.
Given a feasible solution of $\I_{\mathrm{ARP}}$, each airport serves its own unit locally. Each edge of its tree carries at most $K-1$ units toward the airport and requires only one cable. Thus we have $\OPT_{\mathrm{HCCFL}}\le\OPT_{\mathrm{ARP}}$.

We then apply Theorem~\ref{thm:binary} to $\I_{\mathrm{HCCFL}}$ to obtain a cable solution with opening cost $\Phi$ and cable cost $C$.
Close facilities with zero receipt and decompose the flow into unit paths. We may assume that each open airport serves its own unit locally. Otherwise, let $a\leadsto r$ be a path sending a unit from client $a$ to airport $r$, and let $r\leadsto b$ be the path sending the unit of $r$ to airport $b$. We serve the unit of $r$ locally and send the unit of $a$ along $a\leadsto r\leadsto b$. This preserves all receipts, other local assignments, and the flow on every edge. Canceling any cycles does not increase edge loads.
By removing those local units and their receipts, we obtain a flow for which each open airport has remaining capacity $b_r=K-1$ and whose cable cost is at most $C$.

We then apply Lemma~\ref{lem:fleetdecode} to this flow with cable capacity $K-1$ and the open airports as depots. Since each airport's remaining capacity equals the cable capacity, the lemma yields at most one tree at each airport and total edge cost at most $3C$. Each remaining city has one unit of demand and is not an open airport. By integrality, it is served by exactly one tree. Each tree contains only its airport and the cities it serves, so the trees are node-disjoint and each contains at most $K-1$ cities besides its airport.

Finally, we restore the locally served units, adding singleton trees where needed, to obtain a partition of all cities into trees, each containing exactly one open airport. The singleton trees have zero edge cost, and the opening costs of all airports are included in $\Phi$. Thus the total cost is at most $\Phi+3C\le3\Gamma\OPT_{\mathrm{ARP}}$.
\end{proof}

\subsection{Multiple-Depot Split Delivery Vehicle Routing}\label{sec:mdsd}
In the MDSDVRP, let $G=(V,E)$ be a complete graph with metric edge costs $c:E\to\mathbb{Q}_{\ge0}$. The sets $J,U\subseteq V$ contain the clients and depots, respectively. Each client $v\in J$ has demand $d_v\in\mathbb{Z}_{\ge0}$ and each depot $r\in U$ has $m_r\in\mathbb{Z}_{\ge0}$ vehicles of common capacity $Q\in\mathbb{N}^+$. Each route starts and ends at its own depot, so at most $m_r$ routes use depot $r$, and each route delivers at most $Q$ units.
Demand can split among vehicles and depots. The objective is to minimize the total route cost.
Note that demands, fleet sizes, and $Q$ are given in binary and may be superpolynomial in $\size{V}$. As with tree records, we use route records to represent our solution in polynomial size. Each route record specifies the depot, order of visits, delivery amounts, and a binary repetition count.

The MDSDVRP is a well-studied vehicle routing problem. \citet{thesis:hu:SFU:09} solved the problem exactly on paths and obtained a $2$-approximation on trees. The latter gives an $\O(\log\size{V})$-approximation on general metrics by randomized tree embeddings~\citep{journal:fakcharoenphol:JCSS:04}.
For a fixed number $k\ge2$ of depots, \citet{journal:lai:IJOC:23} gave a $(6-\frac{4}{k})$-approximation in time $\O(\size{V}^{k-1}\size{E}^{2(k-1)}\max\{\size{V}^3,(\sum_{r\in U}m_r)^3\})$.

Recently, \citet{preprint:zhao:arxiv:26a} used a polynomial-time $\alpha$-approximation for the traveling salesman problem (TSP) to obtain a $(2\alpha+1+\varepsilon)$-approximation in time $2^{\O(\frac{k\log k}{\varepsilon})}\poly(\size{V})$.
They also obtained a polynomial-time $5$-approximation when $\sum_{r\in U}m_rQ-\sum_{v\in J}d_v=\O(1)$.
Moreover, with capacity $(1+\varepsilon)Q$ per vehicle, they proposed a polynomial-time algorithm with cost at most $\O(\frac{1}{\varepsilon})$ times the optimum under capacity $Q$.

\begin{theorem}\label{thm:mdsd}
The MDSDVRP admits a polynomial-time $6\Gamma$-approximation.
\end{theorem}
\begin{proof}
We keep all nodes and take the clients with positive demand as the clients of the cable instance. For each depot $r$, set $b_r=m_rQ$. We place a free facility of capacity $b_r$ at each depot with $m_r>0$, and set the cable capacity to $Q$.

Consider any feasible MDSDVRP solution. Its routes serve integer demands with vehicle capacity $Q$ and total delivery at most $b_r$ at depot $r$. We can choose integral deliveries on these routes. We first construct a flow network with a start node and a terminal, and separate nodes for clients, routes, and depots. The arc from the start node to client $v$ has capacity $d_v$, and an arc from $v$ to each route visiting it also has capacity $d_v$. Each route has an arc to its depot of capacity $Q$, and each depot $r$ has an arc to the terminal of capacity $b_r$. The original deliveries give a flow of value $\sum_{v\in J}d_v$. Since all capacities are integers, a maximum flow of this value can be chosen integral \citep{book:schrijver:springer:03}. We then take a spanning tree of the nodes and edges used by each route and send its deliveries to the depot along that tree. One cable on each tree edge suffices because the total load on the route is at most $Q$. The sum of these cable flows has cost at most the total route cost by \eqref{eq:subadd} and receives at most $b_r$ units at each depot. Thus $\OPT_{\mathrm{HCCFL}}\le\OPT_{\mathrm{MDSDVRP}}$.

We apply Theorem~\ref{thm:binary} and let $C$ be the cost of its cable solution. We then apply Lemma~\ref{lem:fleetdecode} with the original vehicle capacity $Q$ and facility capacities $b_r=m_rQ$. It gives at most $m_r$ trees at each $r$, each with load at most $Q$; since $Q$ divides each $b_r$, their total cost is at most $3C$. We double each tree and shortcut the resulting walk to a closed route on the depot and the delivery nodes. The routes inherit the tree loads and repetition counts. Their total cost is at most $6C\le6\Gamma\OPT_{\mathrm{MDSDVRP}}$.
\end{proof}

\subsection{Capacitated Location Routing with Hard Facility Capacities}\label{sec:clr}
In the H-CLR, let $G=(V,E)$ be a complete graph with metric edge costs $c:E\to\mathbb{Q}_{\ge0}$. The sets $J,U\subseteq V$ contain the clients and candidate depots, respectively. Each client $v\in J$ has demand $d_v\in\mathbb{Z}_{\ge0}$, and each depot $r\in U$ has capacity $b_r\in\mathbb{N}^+$ and opening cost $\phi_r\in\mathbb{Q}_{\ge0}$. Each route starts and ends at an open depot and delivers at most $Q\in\mathbb{N}^+$ units. Demand can split among routes and depots, and there is no bound on the number of routes. The objective is to minimize the total opening and route cost. We assume $b_r\ge Q$ for every depot.
As in the MDSDVRP, demands and capacities are given in binary, and we use route records to represent our solution.

For the H-CLR with $b_r=\infty$, \citet{journal:harks:TS:13} developed a $4.38$-approximation algorithm, where the demands are unsplit if $d_v\le Q$.
\citet{journal:zhao:IC:26} improved this bound to $4.169$ and obtained a $4.091$-approximation for split demands.
For the H-CLR, \citet{journal:carrasco:EJOR:23} obtained cost at most $(4+\frac{2\rho_{\mathrm{CFL}}}{\varepsilon})$ times the optimum under the original capacities, with depot capacities $b_r+\varepsilon Q$. Here $\rho_{\mathrm{CFL}}$ is an approximation for the CFL (we can take $\rho_{\mathrm{CFL}}=5+\varepsilon'$ for any fixed $\varepsilon'>0$ \citep{conf:bansal:ESA:12}).

\begin{theorem}\label{thm:clr}
When $b_r\ge Q$, the H-CLR admits a polynomial-time $20\Gamma$-approximation.
\end{theorem}
\begin{proof}
We keep all nodes and take the clients with positive demand as the clients of the cable instance. For each depot $r$, we place a facility of capacity $b_r$ and opening cost $\phi_r$ at $r$, and set the cable capacity to $Q$.

Consider any feasible H-CLR solution. As in the proof of Theorem~\ref{thm:mdsd}, we can convert its routes into a cable solution of no greater cost, keeping the same open depots. Thus $\OPT_{\mathrm{HCCFL}}\le\OPT_{\mathrm{HCLR}}$.

We apply Theorem~\ref{thm:binary} and let $\Phi$ and $C$ be the opening and cable costs of its solution, respectively. We then apply Lemma~\ref{lem:fleetdecode} with the original vehicle capacity $Q$ and depot capacities $b_r$. It gives trees of total cost at most $10C$, each with load at most $Q$ and with total receipt at each depot $r$ at most $b_r$. We double each tree and shortcut the resulting walk to a closed route on the depot and the delivery nodes. The routes inherit the tree loads and repetition counts. Their total opening and route cost is at most $\Phi+20C\le20\Gamma\OPT_{\mathrm{HCLR}}$.
\end{proof}

\begin{remark}
The ratio $\frac{\OPT_{\mathrm{HCLR}}}{\OPT_{\mathrm{HCCFL}}}$ can be arbitrarily large when depot capacities are unrestricted. Put $Q$ units of demand at a node $s$ and one free depot of capacity one at each of $Q$ other nodes. The edge cost from $s$ to each depot is one, and the edge cost between any two depots is zero. One cable from $s$ to any depot and zero-cost cables among the depots serve all demand at cost one, while in a routing solution every depot must send a vehicle to $s$ and back for a total route cost of at least $2Q$.

Suppose the H-CLR additionally limits each depot $r$ to $m_r\in\mathbb{Z}_{\ge0}$ vehicles, each used on at most one route. This problem generalizes the MDSDVRP by allowing depot opening costs and capacities. For each depot with $m_r>0$, set $b'_r=\min\{b_r,m_rQ\}$; depots with $m_r=0$ can be omitted while retaining their nodes. Every feasible routing solution delivers at most $b'_r$ units from depot $r$. If $b_r\ge Q$ for every remaining depot, then $b'_r\ge Q$. Lemma~\ref{lem:fleetdecode} therefore gives at most $\ceil{\frac{b'_r}{Q}}\le m_r$ trees at $r$, each yielding one route. The proof of Theorem~\ref{thm:clr}, applied with capacities $b'_r$, thus also gives a $20\Gamma$-approximation.
\end{remark}

\section{Omitted Proofs}\label{sec:omitted}
\subsection{Proof of Lemma~\ref{lem:compatible}}\label{app:compatible}
\begingroup
\def\thelemma{\ref{lem:compatible}}
\begin{lemma}
Let $f$ be a feasible rational flow from sources with integer demands to a given set of facilities with integer capacities, and let $H$ be a tree connecting all sources and these facilities. There exists a polynomial-time algorithm that constructs a feasible integral cable solution with cable cost at most $\frac{L(f)}{Q}+c(H)$.
\end{lemma}
\endgroup
\begin{proof}
We first find an integral flow $f_0$ to these facilities with $L(f_0)\le L(f)$. Replace each edge by two opposite arcs of cost $c_e$. Join each facility to a common terminal by a zero-cost arc whose capacity is the facility's capacity. Each source $s$ supplies $d_s$, every other original node has zero supply, and the terminal receives $d_J$. The flow $f$ gives a feasible flow in this network. Since demands and capacities are integers, we can find an integral minimum-cost flow and cancel opposite flows to obtain $f_0$.

We next construct any integral flow $h$ on $H$ with the same divergence as $f_0$. Such a flow exists because $H$ connects all sources and these facilities. Thus $z=\frac{f_0-h}{Q}$ has zero divergence at every node. We then round $z$ to an integral flow while keeping its divergence zero.

For each edge $e$, require $\floor{z_e}\le z'_e\le\ceil{z_e}$. Let $a_e=h_e+Q\floor{z_e}$ and $b_e=h_e+Q\ceil{z_e}$ be the flows when $z'_e$ takes the lower and upper bounds, respectively. We assign a weight to each value of $z'_e$ in two cases.
If $z_e$ is an integer, assign $z'_e=z_e$ the weight $c_e\ceil{\frac{|a_e|}{Q}}$.
Otherwise, assign each value of $z'_e\in[\floor{z_e},\ceil{z_e}]$ the weight
\begin{equation}\label{eq:rounding-cost}
 c_e\lrA{(\ceil{z_e}-z'_e)\ceil{\frac{|a_e|}{Q}}
 +(z'_e-\floor{z_e})\ceil{\frac{|b_e|}{Q}}}.
\end{equation}
Let $W(z')$ be the sum of these weights over all edges. Each weight is affine in $z'_e$, so minimizing $W(z')$ subject to the bounds and zero divergence is a minimum-cost flow problem. The bounds are integers, and $z$ is a feasible flow with zero divergence. By integrality of minimum-cost flow, we can find an integral flow $z'$ with zero divergence and minimum total weight, so $W(z')\le W(z)$.

Since $z'$ is integral, $h_e+Qz'_e$ equals $a_e$ or $b_e$ by definition. Moreover, the weight of $z'_e$ is the cable cost of the corresponding flow on $e$. Thus $C_Q(h+Qz')=W(z')\le W(z)$.
The flow $h+Qz'$ has the same divergence as $f_0$, so it serves the demands with the facility receipts of $f_0$. To bound its cable cost, it suffices to bound $W(z)$.

First consider an edge $e$ outside $H$. We have $h_e=0$ by definition and $f_{0,e}=Qz_e$ since $z=\frac{f_0-h}{Q}$.
If $z_e$ is an integer, its assigned weight is $c_e|z_e|$. Otherwise, by \eqref{eq:rounding-cost}, its assigned weight is $c_e[(\ceil{z_e}-z_e)|\floor{z_e}|+(z_e-\floor{z_e})|\ceil{z_e}|]=c_e|z_e|.$
In either case, the weight assigned to $z_e$ is $c_e\frac{|f_{0,e}|}{Q}$.

Now consider an edge $e$ of $H$, and let $\theta=z_e-\floor{z_e}$. The weight assigned to $z_e$ is $c_e((1-\theta)\ceil{\frac{|a_e|}{Q}}+\theta\ceil{\frac{|b_e|}{Q}})$, including when $z_e$ is an integer. We have $b_e-a_e\le Q$ and $(1-\theta)a_e+\theta b_e=h_e+Qz_e=f_{0,e}$.
If $a_e\ge 0$ or $b_e\le 0$, then $(1-\theta)|a_e|+\theta|b_e|=|f_{0,e}|$. Using $\ceil{x}\le x+1$, the weight is at most $c_e(\frac{(1-\theta)|a_e|+\theta|b_e|}{Q}+1)=c_e(\frac{|f_{0,e}|}{Q}+1)$.
Otherwise, we have $a_e<0<b_e$, so $|a_e|+|b_e|=b_e-a_e\le Q$. Then $\ceil{\frac{|a_e|}{Q}}=\ceil{\frac{|b_e|}{Q}}=1$, and the weight is $c_e((1-\theta)+\theta)=c_e\le c_e(\frac{|f_{0,e}|}{Q}+1)$.
Thus the weight is at most $c_e(\frac{|f_{0,e}|}{Q}+1)$ in all cases.

Combining the bounds for edges of $H$ and outside $H$ gives $W(z)\le\frac{L(f_0)}{Q}+c(H)\le\frac{L(f)}{Q}+c(H)$.
All steps take polynomial time. Thus the lemma holds.
\end{proof}

\subsection{Proof of Lemma~\ref{lem:residual}}\label{app:residual}
\begingroup
\def\thelemma{\ref{lem:residual}}
\begin{lemma}[Residual assignment]
For a candidate $(X,Y,z)$ satisfying \eqref{eq:aggregate}--\eqref{eq:witnessbudget}, a polynomial-time algorithm either returns a polynomial-size inequality valid for $\mathcal P_B$ and violated by $(X,Y,z)$, or finds: (1) a nonnegative partial assignment $\widetilde X$ to $I$ respecting the capacities $\bar b_i$ with $\ell(\widetilde X)\le4\ell$, and (2) a set of selected group and nonfree facilities of $\I_{\mathrm{aux}}$ and a nonnegative assignment $x$ to them respecting their capacities $b_j$.
The demand of $\I_{\mathrm{aux}}$ at source $s$ is $D_s:=d_s-\sum_{i\in I}\widetilde X_{i,s}$. The assignment $x$ serves nonnegative demands $d'_s\le D_s$ with $\sum_{s\in J}(D_s-d'_s)\le8Q$. The opening cost $O'$ of these selected facilities and the scaled assignment cost $L'$ of $x$ satisfy
\begin{equation}\tag{\ref{eq:aux-cost}}
 O'+L'\le36\left(4\sum_{i\notin I,\,i\text{ nonfree}}\phi_iY_i+\Lambda\right),
\end{equation}
where $\Lambda=42\Omega+44\tau+8\ell+\ell(\widetilde X)\le42\Omega+44\tau+12\ell$.
\end{lemma}
\endgroup
\begin{proof}
We first compute a maximum flow to construct the partial assignment $\widetilde X$ to $I$. For the remaining demand $D_s$, we construct a CFL instance using $\I_{\mathrm{aux}}$ and additional facilities, and seek a fractional assignment with suitable bounds on its receipts and cost. If no such assignment exists, we return an inequality valid for $\mathcal P_B$ and violated by $(X,Y,z)$; otherwise, we round a minimum-cost assignment and remove the additional facilities to obtain $x$.

\paragraph{Partial assignment.}
Our construction is based on the result of \citet{journal:an:SICOMP:17}.
Add an auxiliary start node and terminal. Connect the start node to each source $s$ with capacity $d_s$, each source $s$ to each entry $i\in I$ with capacity $4X_{i,s}$, and each entry $i$ to the terminal with capacity $\bar b_i$. Sources and entries are separate nodes in this network, even when they share a location.
Since $X\ge0$ by \eqref{eq:aggregate}, all arc capacities are nonnegative.
We then compute a maximum flow $M$~\citep{book:schrijver:springer:03}.

To identify the sources whose demands may remain unassigned, consider the residual network of $M$: an arc has a forward residual arc if it has unused capacity and a reverse residual arc if it carries positive flow. Let $J'\subseteq J$ and $I'\subseteq I$ be the sources and entries reachable from the start node in this network.
Remove from $M$ the assignments from $J'$ to $I\setminus I'$, and denote the resulting partial assignment by $\widetilde X$. For $i\notin I$, define $\widetilde X_{i,s}=0$. The remaining demand at source $s$ is $D_s=d_s-\sum_i\widetilde X_{i,s}$.

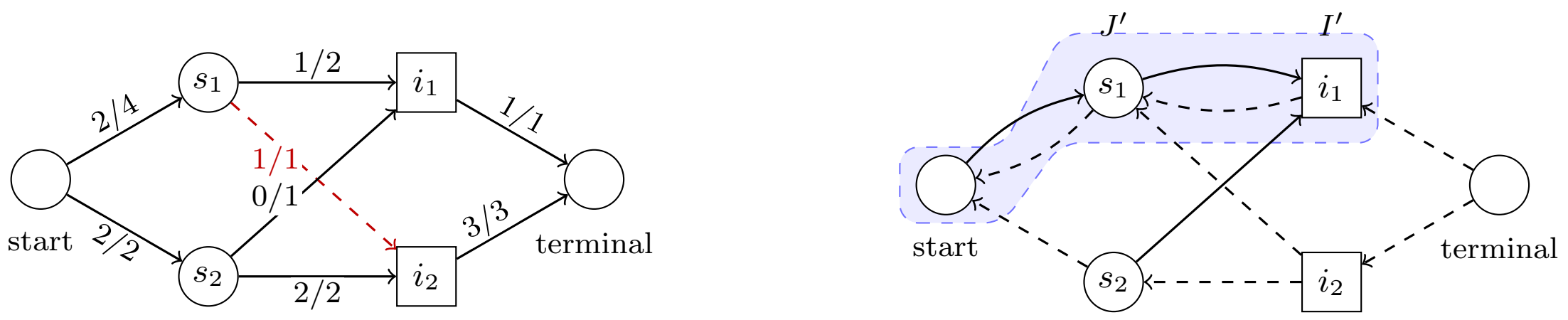
\begin{figure}[!t]
\centering
\begin{subfigure}[t]{0.49\textwidth}
\centering
\begin{tikzpicture}[
  x=0.78cm,y=0.82cm,
  every node/.style={font=\footnotesize},
  source/.style={circle,draw,fill=white,minimum size=5.5mm,inner sep=1pt},
  entry/.style={rectangle,draw,fill=white,minimum size=5.5mm,inner sep=1pt},
  edge/.style={->,line width=0.6pt},
  value/.style={font=\scriptsize,fill=white,inner sep=0.6pt}
]
  \node[source] (a) at (0,0) {};
  \node[source] (s1) at (2,1.1) {$s_1$};
  \node[source] (s2) at (2,-1.1) {$s_2$};
  \node[entry] (i1) at (4.6,1.1) {$i_1$};
  \node[entry] (i2) at (4.6,-1.1) {$i_2$};
  \node[source] (t) at (6.6,0) {};
  \node[below=2pt of a,font=\scriptsize] {start};
  \node[below=2pt of t,font=\scriptsize] {terminal};
  \draw[edge] (a) -- node[value,above,sloped] {$2/4$} (s1);
  \draw[edge] (a) -- node[value,below,sloped] {$2/2$} (s2);
  \draw[edge] (s1) -- node[value,above] {$1/2$} (i1);
  \draw[edge,red!75!black,dashed] (s1) -- node[value,pos=0.27,below] {$1/1$} (i2);
  \draw[edge] (s2) -- node[value,pos=0.27,above] {$0/1$} (i1);
  \draw[edge] (s2) -- node[value,below] {$2/2$} (i2);
  \draw[edge] (i1) -- node[value,above,sloped] {$1/1$} (t);
  \draw[edge] (i2) -- node[value,pos=0.35,above,sloped] {$3/3$} (t);
\end{tikzpicture}
\caption{Maximum flow $M$.}
\end{subfigure}\hfill
\begin{subfigure}[t]{0.49\textwidth}
\centering
\begin{tikzpicture}[
  x=0.78cm,y=0.82cm,
  every node/.style={font=\footnotesize},
  source/.style={circle,draw,fill=white,minimum size=5.5mm,inner sep=1pt},
  entry/.style={rectangle,draw,fill=white,minimum size=5.5mm,inner sep=1pt},
  forward/.style={->,line width=0.6pt},
  reverse/.style={->,dashed,line width=0.6pt}
]
  \filldraw[fill=blue!8,draw=blue!55,dashed,rounded corners=5pt]
    (-0.55,-0.43)--(0.68,-0.43)--(1.38,0.48)--(5.15,0.48)
    --(5.15,1.72)--(1.38,1.72)--(0.68,0.43)--(-0.55,0.43)--cycle;
  \node[source] (a) at (0,0) {};
  \node[source] (s1) at (2,1.1) {$s_1$};
  \node[source] (s2) at (2,-1.1) {$s_2$};
  \node[entry] (i1) at (4.6,1.1) {$i_1$};
  \node[entry] (i2) at (4.6,-1.1) {$i_2$};
  \node[source] (t) at (6.6,0) {};
  \node[below=2pt of a,font=\scriptsize] {start};
  \node[below=2pt of t,font=\scriptsize] {terminal};
  \node[above=2pt of s1,font=\scriptsize] {$J'$};
  \node[above=2pt of i1,font=\scriptsize] {$I'$};
  \draw[forward,bend left=17] (a) to (s1);
  \draw[reverse,bend left=17] (s1) to (a);
  \draw[reverse] (s2) -- (a);
  \draw[forward,bend left=17] (s1) to (i1);
  \draw[reverse,bend left=17] (i1) to (s1);
  \draw[reverse] (i2) -- (s1);
  \draw[forward] (s2) -- (i1);
  \draw[reverse] (i2) -- (s2);
  \draw[reverse] (t) -- (i1);
  \draw[reverse] (t) -- (i2);
\end{tikzpicture}
\caption{Residual network of $M$.}
\end{subfigure}
\caption{The partial-assignment network. In (a), labels show flow/capacity and the red arc's flow is removed. In (b), dashed arrows reverse original arcs and shaded nodes are reachable.}
\label{fig:partial-assignment}
\end{figure}

Since $M$ is a maximum flow, the terminal is not reachable from the start node in its residual network. This gives the following properties.
\begin{enumerate}
 \item Every source in $J\setminus J'$ is fully assigned. Otherwise, the arc from the start node to this source would have unused capacity, making the source reachable.
 \item Every entry in $I'$ is saturated. Otherwise, the terminal would be reachable from the start node, contradicting the optimality of $M$.
 \item No assignment from $J\setminus J'$ goes to $I'$. Otherwise, a positive assignment would make its source reachable through a reverse residual arc.
\end{enumerate}
An illustration of these properties is shown in Figure~\ref{fig:partial-assignment}.
These properties remain valid after we remove the assignments above. Moreover, each arc from $J'$ to $I\setminus I'$ is saturated in $M$; otherwise its endpoint would be reachable. Each such arc carries $4X_{i,s}$ units in $M$, which are removed. Thus the remaining demand satisfies
\begin{equation}\label{eq:direct-high}
 D_s\ge4\sum_{i\in I\setminus I'}X_{i,s},\qquad s\in J'.
\end{equation}
Moreover, $0\le\widetilde X_{i,s}\le4X_{i,s}$ and $\widetilde X$ respects the capacities, so its scaled cost is $\ell(\widetilde X)\le4\ell$.

\paragraph{Assigning the remaining demand.}
For comparison with points of $\mathcal P_B$, we fix $I$, $I'$, $J'$, the groups $\G$ and sets $R_i$ from Section~\ref{sec:ingredients}, and the partial assignment $\widetilde X$ constructed using $(X,Y,z)$.

We construct a CFL instance by extending $\I_{\mathrm{aux}}$ from Section~\ref{sec:residual}, using the remaining demand $D_s$ at each source $s$ and capacity $\frac{b_j}{4}$ for each group or nonfree facility $j$. We seek a nonnegative assignment of all $D_s$ units from each source $s$. We add a facility $u_t$ at each source $t\in J'$, with capacity $\frac{D_t}{4}$, and a facility $u_0$ at a source $s_0\in J$, with capacity $2Q$. We call $u_t$ and $u_0$ the additional facilities and keep them distinct from one another and from the other facilities, with their own capacities, even when they share a location. The facilities $u_t$ allow us to reassign demand through sources, while $u_0$ accounts for demand that may remain unassigned after rounding.
The opening cost of each nonfree facility $i\notin I$ is $\phi_i$, and every other facility has zero opening cost. We use the scaled assignment costs $\hat c=\frac{c}{Q}$ in the extended metric of Section~\ref{sec:residual}. For a fractional solution, we use the opening fractions $y_j$ from that section for the facilities of $\I_{\mathrm{aux}}$ and opening fraction one for $u_t$ and $u_0$.

We next impose bounds on the amounts assigned to these facilities.
The following table gives the bounds on each facility's total receipt and its receipt from $s$, as well as its scaled cost per unit.
\begin{equation}\label{eq:terminal}
 \begin{array}{c|c|c|c}
  \text{facility} & \text{receipt bound} & \text{bound from $s$} & \text{unit cost}\\\hline
  i\text{ nonfree},\ i\notin I & b_iY_i & D_sY_i & \hat c(i,s)\\
  \G & 3\sum_{i\in\G}\bar b_i z(\partial R_i) & D_s & \hat c(s,\G)\\
  u_t\ (t\in J') & \sum_{i\in I\setminus I'}X_{i,t} & D_s & \hat c(s,t)\\
  u_0 & 2Q & D_s & \hat c(s,s_0)
 \end{array}
\end{equation}
We also require that the total scaled assignment cost be at most $\Lambda=42\Omega+44\tau+8\ell+\ell(\widetilde X)$. At $(X,Y,z)$, $\ell(\widetilde X)\le4\ell$, so $\Lambda\le42\Omega+44\tau+12\ell$.

\paragraph{Choosing facilities.}
Suppose there is an assignment satisfying \eqref{eq:terminal} at $(X,Y,z)$ with scaled cost at most $\Lambda$. We first use a minimum-cost flow to find such a rational assignment $\widehat x$, where $\widehat x_{j,s}$ is the amount assigned from source $s$ to facility $j$.
Then we use $\widehat x$ to construct the assignment $x$. If no such assignment exists, we return an inequality valid for $\mathcal P_B$ and violated by $(X,Y,z)$, as shown at the end of the proof.

We first verify that $\widehat x$ and the opening fractions form a feasible fractional solution to the CFL relaxation. The assignment $\widehat x$ serves all $D_s$, and the opening fractions lie in $[0,1]$. By \eqref{eq:terminal}, at a nonfree facility $i\notin I$, the total receipt is at most $b_iY_i=\frac{b_i}{4}y_i$, and the receipt from $s$ is at most $D_sY_i\le D_sy_i$. A group facility $\G$ receives at most $3\sum_{i\in\G}\bar b_i z(\partial R_i)\le \frac{b_{\G}}{4}$ by \eqref{eq:small-cut} with $\delta=\frac{1}{12}$. The receipt at $u_t$ is at most $\frac{D_t}{4}$ by \eqref{eq:direct-high}, and the receipt at facility $u_0$ is at most $2Q$. The group and additional facilities have opening fraction one, so their receipt from any source $s$ is bounded by $D_s$. Thus any nonnegative assignment that serves all $D_s$ and satisfies \eqref{eq:terminal} at $(X,Y,z)$, together with these opening fractions, is a feasible fractional solution to the CFL relaxation.

We next round $\widehat x$ using the following form of the result of \citet{report:abrams:CMU:02} for rational demands and capacities.
\begin{claim}[{\citet[Theorem~14]{journal:an:SICOMP:17}}]\label{clm:cfl-rounding}
Given a feasible fractional solution to the CFL relaxation with rational demands and capacities, there exists a polynomial-time algorithm that selects facilities and assigns all demand to them at total opening and assignment cost at most $18$ times the fractional cost.
Moreover, each selected facility receives at most twice its capacity.
\end{claim}

Omit any facility of zero capacity and any source with $D_s=0$ (if all $D_s=0$, return the empty assignment with $O'=L'=0$).
We then apply Claim~\ref{clm:cfl-rounding} to $\widehat x$ and the opening fractions with the scaled costs $\hat c$ in the extended metric of Section~\ref{sec:residual}. The fractional opening cost is $4\sum_{i\notin I,\,i\text{ nonfree}}\phi_iY_i$, and the scaled assignment cost is at most $\Lambda$. Thus the opening cost $O_1$ and scaled assignment cost $L_1$ after rounding satisfy
\begin{equation}\label{eq:cfl-rounding-cost}
 O_1+L_1\le18\left(4\sum_{i\notin I,\,i\text{ nonfree}}\phi_iY_i+\Lambda\right).
\end{equation}
Each group or nonfree facility $j$ receives at most $\frac{b_j}{2}$, each $u_t$ at most $\frac{D_t}{2}$, and facility $u_0$ at most $4Q$.

\paragraph{Removing the additional facilities.}
To remove the facilities $u_t$, we first replace each rounded assignment from $s$ to $u_t$ by flow from source $s$ to source $t$ at the same scaled cost per unit $\hat c(s,t)$.
We keep the assignments to group facilities, nonfree facilities outside $I$, and $u_0$ as arcs to separate facility nodes, even when they share a location with a source.
Let $q_{u_s}^{(1)}$ be the receipt at $u_s$ after rounding, with $q_{u_s}^{(1)}=0$ if $u_s$ is absent. The net outflow at source $s$ is $D_s-q_{u_s}^{(1)}\ge \frac{D_s}{2}$.

We then decompose this flow into paths and cycles \citep{book:ahuja:prentice:93}. The paths starting at each source $s$ carry $D_s-q_{u_s}^{(1)}$ units and end at group and nonfree facilities or facility $u_0$. Retain $\frac{D_s}{2}$ units from these paths, splitting a path amount if needed. Double the retained amounts and assign them directly from each source to the facilities at the ends of the paths. By the triangle inequality, this serves all $D_s$ units at each source at scaled cost at most $2L_1$. Each facility receives at most twice its rounded receipt, so the receipt at a group or nonfree facility $j$ is at most $b_j$ and the receipt at facility $u_0$ is at most $8Q$.

Finally, we delete the assignments to facility $u_0$, let $x$ be the resulting assignment, and select the group and nonfree facilities of $\I_{\mathrm{aux}}$ that receive positive demand in $x$.
The amounts $d'_s$ served by $x$ satisfy $0\le d'_s\le D_s$ and $\sum_{s\in J}(D_s-d'_s)\le8Q$. The opening cost is $O'\le O_1$ and the scaled assignment cost is $L'\le2L_1$, so $O'+L'\le2(O_1+L_1)$. By \eqref{eq:cfl-rounding-cost}, this implies \eqref{eq:aux-cost}. All these steps take polynomial time.

It remains to show how to find an assignment satisfying \eqref{eq:terminal} at $(X,Y,z)$ with scaled cost at most $\Lambda$, or return an inequality valid for $\mathcal P_B$ and violated by $(X,Y,z)$.
\begin{claim}\label{clm:transport-feasible}
For every point $(X',Y',z')\in\mathcal P_B$, there is a nonnegative assignment that serves $D_s$ at each source, satisfies \eqref{eq:terminal}, and has scaled cost at most $\Lambda$, where the bounds in \eqref{eq:terminal} and $\Lambda$ are evaluated at $(X',Y',z')$, while $\widetilde X$ and $D_s$ remain fixed.
\end{claim}
\begin{claimproof}
We keep $I,I',J'$, the groups $\G$ and their sources $s_{\G}$ and costs $c_{\G}$, the sets $R_i$, $\widetilde X$, and $D_s$ fixed as above.

Let $(X^0,Y^0,z^0)$ be an arbitrary integer point defining $\mathcal P_B$.
As in the proof of Lemma~\ref{lem:residual-flow}, we first use the difference $X^0-\widetilde X$ to construct paths. Keep sources and entries as separate nodes. For each pair $(i,s)$, direct a positive difference from $s$ to $i$ and a negative difference from $i$ to $s$, using its absolute value as the flow on the arc and $\hat c(i,s)$ as the cost per unit. The net outflow at source $s$ is $\sum_i(X^0_{i,s}-\widetilde X_{i,s})=D_s$ by \eqref{eq:aggregate}. Decompose the flow into cycles and paths from nodes with net outflow to nodes with net inflow. We keep the paths starting at sources, which carry $D_s$ units from each $s$. Their total scaled cost is at most $\ell(X^0)+\ell(\widetilde X)$.

We stop each path at its first entry in $I\setminus I'$, if it visits one. By the first property of $\widetilde X$ above, each path starts in $J'$. Entries outside $I$ have no arcs back to sources, since $\widetilde X$ is zero there. Thus each intermediate entry lies in $I'$, and the third property ensures that each arc out of $I'$ returns to a source in $J'$. By the second property, each entry $i\in I'$ receives $\bar b_i$ under $\widetilde X$. Its receipt under $X^0$ is at most $\bar b_i$ by \eqref{eq:free-capacity}--\eqref{eq:free-edge-bounds}, so its net inflow in the difference is nonpositive and no path ends there.

First consider a path from $s$ that stops at an entry in $I\setminus I'$. Its last source $t$ belongs to $J'$, so we assign the path amount directly from $s$ to $u_t$. The total flow on the last arcs from $t$ to entries $i\in I\setminus I'$ is at most $\sum_{i\in I\setminus I'}X^0_{i,t}$, as required by \eqref{eq:terminal}. By the triangle inequality, the direct assignment to $u_t$ costs at most the part of the path from $s$ to $t$.

Next, we replace each path ending at a nonfree facility $i\notin I$ by a direct assignment from its starting source. By \eqref{eq:nonfree-capacity}, the total receipt at $i$ is at most $b_iY_i^0$. Since $(X^0,Y^0,z^0)$ is an integer point, $Y_i^0\in\{0,1\}$ by \eqref{eq:free-edge-bounds}. The receipt from $s$ is zero if $Y_i^0=0$ and at most $D_s$ if $Y_i^0=1$, so it is at most $D_sY_i^0$. The triangle inequality bounds the cost of each direct assignment by the cost of its path.

The remaining paths end at entries of $I_{\rm sm}$, by the definition of $I$ in Section~\ref{sec:residual}. Each such entry has positive receipt under $X^0$, since $\widetilde X$ is zero outside $I$. By the definition of $\mathcal P_B$, edges with positive $z^0$-value connect each of these entries to a source. Together with $\T$, these edges contain a tree connecting the entries and all sources, of cost at most $\Omega^0+\tau$, where $\Omega^0=\sum_e c_ez_e^0$. Here we use that $z_e^0\in\{0,1\}$ by \eqref{eq:free-edge-bounds}.

We traverse each edge of this tree twice and list the entries in the order in which they are first visited. Divide the list into consecutive sets, ending each set as soon as its total capacity reaches $2Q$. Since each entry has capacity less than $Q$, each complete set has capacity in $[2Q,3Q)$. The entries left over have total capacity less than $2Q$. We assign the amounts of the paths ending at these entries directly from their starting sources to facility $u_0$. By \eqref{eq:free-capacity}, these amounts sum to less than $2Q$. Since $\T$ connects all sources, each such unit costs at most $\frac{\tau}{Q}$, for a total scaled cost of at most $2\tau$.

For each complete set $S$, let $\T_S$ be the subtree connecting its entries. The part of the traversal used to list $S$ visits all its entries and therefore uses each edge of $\T_S$. These parts are disjoint intervals of the traversal, so $\sum_S c(\T_S)\le2(\Omega^0+\tau)$. Set $c_S=c(\T_S)+\min_{i\in S}c(i,J)$. A shortest path attaining this minimum, together with $\T_S$, contains a tree of cost at most $c_S$ connecting the entries of $S$ to a source. By Lemma~\ref{lem:cheap-groups}, entries of $S$ in groups $\G$ with $c_{\G}\le3c_S$ have total capacity at least $Q$.

We split the amount of each path ending in $S$ among these entries in proportion to their capacities. If a share goes to entry $j\in S\cap\G$, we assign that share directly from the path's starting source to the group facility $\G$. By \eqref{eq:free-capacity}, the total receipt under $X^0$ in $S$ is less than $3Q$. With at least $Q$ of capacity available for this split, each such $j$ receives at most $3\bar b_j$. Each of these entries had positive receipt under $X^0$, so $z^0(\partial R_j)\ge1$ by the property following \eqref{eq:small-cut}. Since the sets $S$ are disjoint, summing over the entries of $\G$ gives the receipt bound $3\sum_{j\in\G}\bar b_jz^0(\partial R_j)$ in \eqref{eq:terminal}. Each group facility, each $u_t$, and facility $u_0$ receives at most $D_s$ units from $s$, since the paths starting at $s$ carry $D_s$ units in total.

To bound the scaled assignment cost to the group facilities, consider a share of a path from $s$ to $i\in S$ reassigned to $j\in S\cap\G$. We can connect $i$ to $j$ in $\T_S$ and $j$ to $s_{\G}$ in $\T_{\G}$. By the triangle inequality and the definition of $\hat c(s,\G)$ in Section~\ref{sec:residual},
\begin{equation}\label{eq:transport-group-connection}
 Q\hat c(s,\G)=c(s,s_{\G})+c_{\G}
 \le c(s,i)+c(\T_S)+2c_{\G}\le c(s,i)+7c_S.
\end{equation}
The term $c(s,i)$ is at most the sum of the edge costs on the original path. To sum the extra cost $\frac{7c_S}{Q}$ per unit, we bound the path amounts ending in $S$ by the total receipt under $X^0$ in $S$. For the $c(\T_S)$ term, this receipt is less than $3Q$. For the other term, $\min_{j\in S}c(j,J)\le c(i,s)$ for $i\in S$ and $s\in J$. Since the sets $S$ are disjoint,
\begin{equation}\label{eq:transport-group-sum}
 \sum_S\frac{c_S}{Q}\sum_{i\in S,\,s\in J}X^0_{i,s}
 \le 3\sum_S c(\T_S)+\ell(X^0)
 \le 6(\Omega^0+\tau)+\ell(X^0).
\end{equation}
An illustration of the reassignment of paths to group facilities is shown in Figure~\ref{fig:residual-transport}.

\begin{figure}[!t]
\centering
\begin{subfigure}[t]{0.49\textwidth}
\centering
\begin{tikzpicture}[
  x=0.8cm,y=0.8cm,
  every node/.style={font=\scriptsize},
  vertex/.style={circle,fill=black,inner sep=1.35pt},
  patharc/.style={->,line width=0.65pt},
  reversearc/.style={->,dashed,line width=0.65pt}
]
  \node[vertex,label=above:{$s\in J'$}] (s) at (0,2.15) {};
  \node[vertex,label=above:{$u\in I'$}] (u) at (1.35,2.15) {};
  \node[vertex,label=left:{$t\in J'$}] (t) at (1.35,1.05) {};
  \node[vertex] (stop) at (3.05,2.15) {};
  \node[vertex] (nonfree) at (3.05,1.05) {};
  \node[vertex] (small) at (3.05,-0.05) {};
  \draw[patharc] (s) -- (u);
  \draw[reversearc] (u) -- (t);
  \draw[patharc] (t) -- (stop);
  \draw[patharc] (t) -- (nonfree);
  \draw[patharc] (t) -- (small);
  \draw[line width=0.6pt] (stop) circle (3pt);
  \node[anchor=west] at (3.15,2.15) {$I\setminus I'\to u_t$};
  \node[anchor=west] at (3.15,1.05) {nonfree $i\notin I\to i$};
  \node[anchor=west] at (3.15,-0.05) {$I_{\rm sm}\to\G$ or $u_0$};
\end{tikzpicture}
\caption{Path destinations.}
\end{subfigure}\hfill
\begin{subfigure}[t]{0.49\textwidth}
\centering
\begin{tikzpicture}[
  x=0.8cm,y=0.8cm,
  every node/.style={font=\scriptsize},
  vertex/.style={circle,fill=black,inner sep=1.35pt},
  leaf/.style={rectangle,draw,fill=white,minimum size=4.5mm,inner sep=1pt}
]
  \node[vertex,label=above:{$s$}] (s) at (0,2.15) {};
  \node[vertex,label=above:{$s_{\G}$}] (sg) at (3.5,2.15) {};
  \node[leaf,label=right:{$\G$}] (leaf) at (5.1,2.15) {};
  \node[vertex,label=below:{$i\in S$}] (i) at (0.65,0.25) {};
  \node[vertex,label=below:{$j\in S\cap\G$}] (j) at (2.45,0.25) {};
  \draw[densely dashed,line width=0.6pt] (s) -- node[left] {$c(s,i)$} (i);
  \draw[blue!65!black,line width=1pt] (i) -- node[above] {$\T_S$} (j);
  \draw[orange!75!black,line width=1pt] (j) -- node[right] {$\T_{\G}$} (sg);
  \draw[teal!75!black,line width=0.85pt] (s) -- node[above] {$c(s,s_{\G})$} (sg);
  \draw[teal!75!black,line width=0.85pt] (sg) -- node[above] {$c_{\G}$} (leaf);
\end{tikzpicture}
\caption{Assignment to a group.}
\end{subfigure}
\caption{Path destinations and assignment to a group facility. In (a), dashed arrows denote reverse arcs.}
\label{fig:residual-transport}
\end{figure}
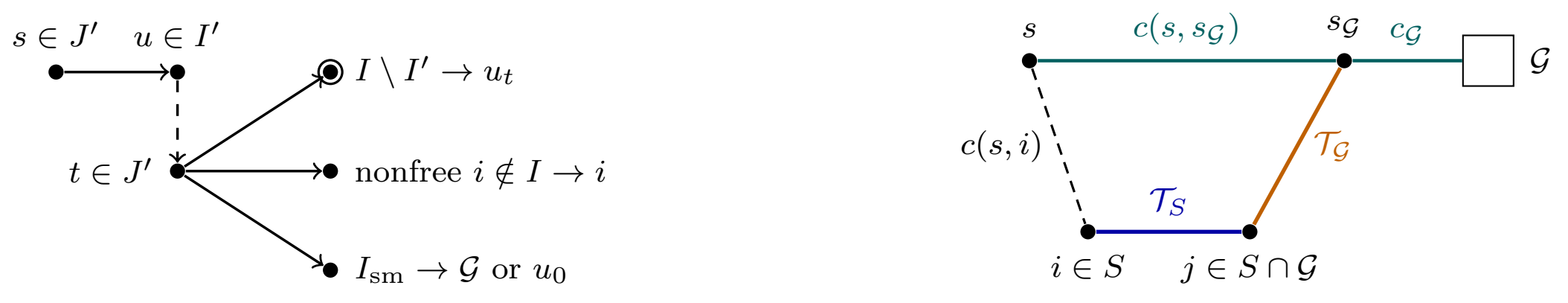

By \eqref{eq:transport-group-connection} and \eqref{eq:transport-group-sum}, the extra scaled cost of the assignments to group facilities is at most $42(\Omega^0+\tau)+7\ell(X^0)$. Adding the original path cost of at most $\ell(X^0)+\ell(\widetilde X)$ and the cost of at most $2\tau$ for facility $u_0$, we obtain $42\Omega^0+44\tau+8\ell(X^0)+\ell(\widetilde X)$. This is the bound $\Lambda$ evaluated at $(X^0,Y^0,z^0)$. Since the integer point was arbitrary, we obtain such an assignment for every integer point defining $\mathcal P_B$. Each point of $\mathcal P_B$ is a convex combination of these integer points. The bounds in \eqref{eq:terminal} and the cost bound $\Lambda$ are affine in $(X,Y,z)$, so using the same coefficients to combine the corresponding assignments gives an assignment satisfying these bounds at that point.
\end{claimproof}

\paragraph{Testing the candidate.}
If no assignment satisfying \eqref{eq:terminal} exists at $(X,Y,z)$, or if the minimum cost is greater than $\Lambda$, we can find in polynomial time a polynomial-size inequality that $(X,Y,z)$ violates and that is satisfied by every point admitting an assignment satisfying \eqref{eq:terminal} with cost at most its corresponding bound $\Lambda$ \citep{book:grotschel:springer:93}. By Claim~\ref{clm:transport-feasible}, the inequality is valid for $\mathcal P_B$.
\end{proof}

\subsection{Proof of Lemma~\ref{lem:fleetdecode}}\label{app:fleetdecode}
\begingroup
\def\thelemma{\ref{lem:fleetdecode}}
\begin{lemma}
Let $f$ be an integral cable flow from clients $J$ to depots $U$ with cable capacity $Q$, and let $q_r$ be the receipt at depot $r$. Suppose each depot $r$ with $q_r>0$ has an integer capacity $b_r\ge Q$ with $q_r\le b_r$. A polynomial-time algorithm serves all demand with trees of total cost at most $10C_Q(f)$, stored in $\O(nm)$ records. At most $\ceil{\frac{b_r}{Q}}$ trees are rooted at each receiving depot $r$, and none at other depots. Each tree serves at most $Q$ units in integer amounts at its clients and contains only its depot and the nodes it serves; the trees at $r$ together serve at most $b_r$ units. Moreover, if $b_r$ is divisible by $Q$ for every receiving depot $r$, the cost bound improves to $3C_Q(f)$.
\end{lemma}
\endgroup
\begin{proof}
We first cancel directed cycles and decompose $f$ into polynomially many paths $\pi$ from clients to receiving depots, including zero-edge paths for local service \citep{book:ahuja:prentice:93}, where path $\pi$ carries $a_\pi\in\mathbb{Z}_{>0}$ units. Let $U^*=\{r\in U:q_r>0\}$ be the set of receiving depots. For each client $v$ and depot $r$, let $x_{v,r}$ be the total amount on the paths from $v$ to $r$. These assignments meet all demands and give each depot its receipt $q_r$. By the triangle inequality, we have
\begin{equation}\label{eq:application-assignment}
 \sum_{v\in J,\,r\in U^*}c(v,r)x_{v,r}\le\sum_\pi a_\pi c(\pi)\le L(f).
\end{equation}

\paragraph{Labeling the units.}
For each $r\in U^*$, we first add $b_r-q_r$ dummy units at $r$. We then assign initial label $r$ to these dummy units and to the $x_{v,r}$ units of demand at each client $v$. Each unit $u$ now has an initial depot label $r(u)\in U^*$ and a location $v(u)\in V$, and each $r\in U^*$ has $b_r$ units in its own component of the flow support.

Take a spanning tree $\T_W$ of each component $W$ of the flow support, including isolated nodes with local service, so $\sum_W c(\T_W)\le C_Q(f)$. Traverse $\T_W$ using each edge twice and order the receiving depots $r\in U^*\cap W$ by their first visits. List the $N_W:=\sum_{r\in U^*\cap W}b_r$ units of $W$ grouped by their depot labels in this order.

Let $N_W=m_WQ+s_W$ with $m_W\in\mathbb{Z}_{\ge0}$ and $0\le s_W<Q$. We partition the list into $m_W$ complete blocks of $Q$ units and a final block $B_W^0$ of $s_W$ units, which is empty when $s_W=0$. Since $b_r\ge Q$ for all $r\in U^*$, each complete block contains units labeled by at most two consecutive depots, and $B_W^0$ contains units of only the last depot $r_W$ in $W$. Set $B_W^0$ aside. For each complete block containing units labeled by two depots, introduce a distinct new label $g$ for its $Q$ units. Let $r_g$ and $h_g$ be these depots, with respective counts $b_{g,1}$ and $b_{g,2}$, so $b_{g,1}+b_{g,2}=Q$. Each resulting label $g$, either an original depot label or a new label, appears on $M_gQ$ units for an integer $M_g\ge1$; for a new label, $M_g=1$.
An illustration of the partition for one component is shown in Figure~\ref{fig:fleet-label-blocks}.
Since each new label $g$ corresponds to a boundary between consecutive depots $r_g,h_g$ in the traversal order, the corresponding portions of the traversal occur without overlap, so
\[
 \sum_{g\text{ new}}c(r_g,h_g)\le 2\sum_W c(\T_W).
\]

\begin{figure}[t]
\centering
\begin{tikzpicture}[font=\small, line width=.65pt, >=stealth]
  \node[above] at (1.5,1.0) {$\T_W$};
  \draw (0,0) -- (1.5,0) -- (3,0);
  \foreach \x in {0,1.5,3} {\fill (\x,0) circle (2pt);}
  \draw[->,gray] (.14,.14) -- (1.35,.14);
  \draw[->,gray] (1.64,.14) -- (2.85,.14);
  \draw[->,gray] (1.35,-.14) -- (.14,-.14);
  \draw[->,gray] (2.85,-.14) -- (1.64,-.14);
  \node[above] at (0,.23) {$b_{r_1}=5$};
  \node[above] at (1.5,.23) {$b_{r_2}=6$};
  \node[above] at (3,.23) {$b_{r_3}=4$};
  \node[below] at (0,-.23) {$r_1$};
  \node[below] at (1.5,-.23) {$r_2$};
  \node[below] at (3,-.23) {$r_3$};
  \begin{scope}[shift={(4.35,0)}]
    \node[above] at (3.12,1.0) {$Q=4$};
    \foreach \i in {0,...,14} {
      \pgfmathtruncatemacro{\lab}{ifthenelse(\i<5,1,ifthenelse(\i<11,2,3))}
      \ifnum\lab=1 \def\shade{black!5}\else
        \ifnum\lab=2 \def\shade{black!20}\else\def\shade{black!35}\fi
      \fi
      \pgfmathsetmacro{\xa}{.38*\i+.18*floor(\i/4)}
      \pgfmathsetmacro{\xb}{\xa+.38}
      \filldraw[fill=\shade,draw=black,line width=.35pt] (\xa,-.1) rectangle (\xb,.55);
      \node[font=\scriptsize] at ({\xa+.19},.225) {$r_{\lab}$};
    }
    \foreach \a/\b in {0/1.52,1.70/3.22,3.40/4.92,5.10/6.24}
      {\draw[line width=.9pt] (\a,-.1) rectangle (\b,.55);}
    \foreach \x in {1.61,3.31,5.01}
      {\draw[densely dashed,line width=.8pt] (\x,-.38) -- (\x,.83);}
    \node[above,inner sep=1pt] at (2.46,.55) {$g$};
    \node[below] at (.76,-.12) {$Q$};
    \node[below] at (2.46,-.12) {$Q$};
    \node[below] at (4.16,-.12) {$Q$};
    \node[below] at (5.67,-.12) {$B_W^0$};
  \end{scope}
\end{tikzpicture}
\caption{A traversal of $\T_W$ from $r_1$ to $r_3$ and back. The boxes show the units in order of their initial depot labels. The second block receives a new label $g$, with $r_g=r_1$, $h_g=r_2$, $b_{g,1}=1$, and $b_{g,2}=3$.}
\label{fig:fleet-label-blocks}
\end{figure}
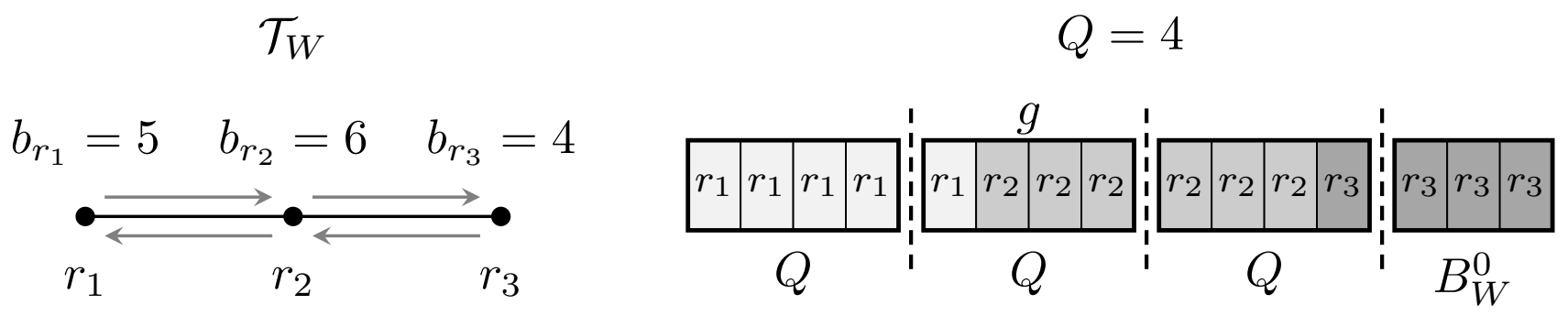

\paragraph{Forming and assigning the paths.}
In each component $W$, list the $m_WQ$ units in the complete blocks by their locations $v(u)$ in the first-visit order of the same traversal of $\T_W$, keeping units with the same initial label and location consecutive. We divide this list into $m_W$ blocks of $Q$ units, and form a path $P=(u_1,\dots,u_Q)$ on the units of each block, where the edge between $u_j$ and $u_{j+1}$ has cost $c(v(u_j),v(u_{j+1}))$. Since the traversal intervals of the blocks are disjoint, $\sum_P c(P)\le 2\sum_W c(\T_W)$.
For each path $P$ and each label $g$ appearing in $P$, let $n_{P,g}$ be the number of units in $P$ with label $g$, set $y_{P,g}=\frac{n_{P,g}}{Q}$, and define
\[
 c_{P,g}=\min_{u\in P:\,\text{label}(u)=g}c(v(u),r(u)).
\]
If $g$ does not appear in $P$, set $n_{P,g}=y_{P,g}=0$. Then $\sum_g y_{P,g}=1$ for each $P$ and $\sum_P y_{P,g}=M_g$ for each $g$. For $y_{P,g}>0$, $c_{P,g}$ is at most the average connection cost of those $n_{P,g}$ units to their original depots. Since dummy units have zero cost to their depots, by \eqref{eq:application-assignment}, we have
\[
 \sum_{P,g:\,y_{P,g}>0}y_{P,g}c_{P,g}
 \le\frac{1}{Q}\sum_{P,g:\,y_{P,g}>0}n_{P,g}c_{P,g}
 \le\frac{1}{Q}\sum_{v\in J,\,r\in U^*}c(v,r)x_{v,r}
 \le\frac{L(f)}{Q}.
\]
By integrality of minimum-cost flow \citep{book:schrijver:springer:03}, we can assign each path $P$ to a label $g(P)$ appearing in $P$ at no greater cost, with $M_g$ paths assigned to each $g$. For each path $P$, choose a unit $u^*\in P$ with label $g(P)$ such that $c(v(u^*),r(u^*))=c_{P,g(P)}$. We consider two cases. If $g(P)$ is a new label $g$, name its two depots $r_g$ and $h_g$ so that $r_g=r(u^*)$, with respective counts $b_{g,1}$ and $b_{g,2}$. Otherwise, $g(P)$ is an original depot label $g$, and we let $r_g=g$. In either case, we attach $P$ by the edge $e_P=r_{g(P)}v(u^*)$, at cost $c_{P,g(P)}$. Then
\[
 \sum_P\bigl(c(P)+c(e_P)\bigr)\le 2\sum_W c(\T_W)+\frac{L(f)}{Q}.
\]

\paragraph{Constructing trees.}
Consider a path $P=(u_1,\dots,u_Q)$ assigned to $g=g(P)$, attached via $e_P=r_gv(u_k)$ for some $1\le k\le Q$. If $g$ represents one depot $r_g$, set $P_1=P$ and $P_2=\varnothing$. Otherwise, $g$ represents $r_g$ and $h_g$ with counts $b_{g,1}+b_{g,2}=Q$, and we split $P$ into $P_1=(u_1,\dots,u_{b_{g,1}})$ for $r_g$ and $P_2=(u_{b_{g,1}+1},\dots,u_Q)$ for $h_g$. For $P_1$, traverse $e_P$ from $r_g$ to $v(u_k)$, follow $P$ to $v(u_1)$, and, if $u_k\in P_1$, traverse $P_1$. For $P_2$ (when $P_2\ne\varnothing$), go from $h_g$ to $r_g$, follow $e_P$ to $v(u_k)$ and $P$ to $v(u_Q)$, and, if $u_k\in P_2$, traverse $P_2$ in reverse. Shortcut each walk to its depot and the clients whose demand is in its part, take a spanning tree, and omit parts with no demand. Each resulting tree serves at most $Q$ units and contains only its depot and the nodes it serves.
By the triangle inequality, shortcutting and taking a spanning tree do not increase cost. The portions of $P$ from $u_k$ to its two endpoints $u_1$ and $u_Q$ partition $P$ and thus have total cost $c(P)$, while $c(P_1)+c(P_2)\le c(P)$. Hence the two trees together cost at most $2c(P)+2c_{P,g}+c(r_g,h_g)$ when $P_2\ne\varnothing$, and at most $2c(P)+c_{P,g}$ when $P_2=\varnothing$.
An illustration of the two walks is shown in Figure~\ref{fig:tree-construction}.

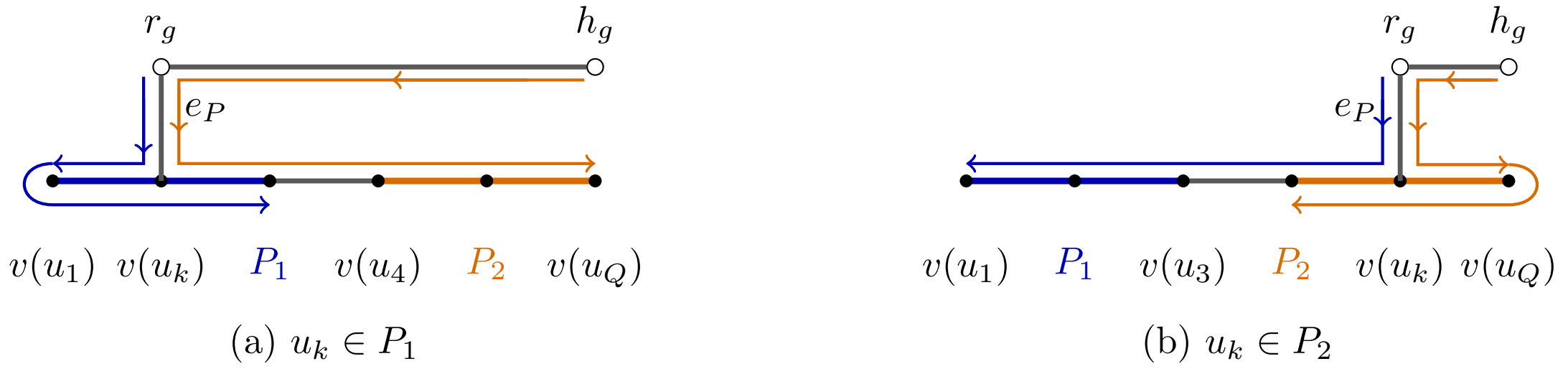
\begin{figure}[!t]
\centering
\begin{subfigure}[t]{0.49\textwidth}
\centering
\begin{tikzpicture}[
  x=1.25cm,
  font=\small,
  unit/.style={circle,fill=black,inner sep=1.2pt},
  depot/.style={circle,draw,fill=white,inner sep=1.5pt},
  first/.style={draw=blue!70!black,line width=0.8pt,->},
  second/.style={draw=orange!85!black,line width=0.8pt,->}
]
  \draw[blue!70!black,line width=1.7pt] (0,0)--(1.6,0);
  \draw[black!65,line width=1.3pt] (1.6,0)--(2.4,0);
  \draw[orange!85!black,line width=1.7pt] (2.4,0)--(4,0);
  \foreach \x in {0,0.8,1.6,2.4,3.2,4} \node[unit] at (\x,0) {};
  \node[depot,label=above:$r_g$] (r) at (0.8,1.05) {};
  \node[depot,label=above:$h_g$] (h) at (4,1.05) {};
  \draw[black!65,line width=1.3pt] (h)--(r)--(0.8,0);
  \node at (1.13,0.65) {$e_P$};
  \draw[first] (0.67,0.96) -- (0.67,0.16) -- (0,0.16);
  \draw[first] (0.67,0.75) -- (0.67,0.25);
  \draw[first] (0,0.16) .. controls (-0.28,0.16) and (-0.28,-0.22) .. (0,-0.22) -- (1.6,-0.22);
  \draw[second] (3.92,0.93) -- (0.93,0.93) -- (0.93,0.16) -- (4,0.16);
  \draw[second] (3.5,0.93) -- (2.5,0.93);
  \draw[second] (0.93,0.85) -- (0.93,0.45);
  \node[below] at (0,-0.48) {$v(u_1)$};
  \node[below] at (0.8,-0.48) {$v(u_k)$};
  \node[below,blue!70!black] at (1.6,-0.48) {$P_1$};
  \node[below] at (2.4,-0.48) {$v(u_4)$};
  \node[below] at (4,-0.48) {$v(u_Q)$};
  \node[below,orange!85!black] at (3.2,-0.48) {$P_2$};
\end{tikzpicture}
\caption{$u_k\in P_1$}
\end{subfigure}\hfill
\begin{subfigure}[t]{0.49\textwidth}
\centering
\begin{tikzpicture}[
  x=1.25cm,
  font=\small,
  unit/.style={circle,fill=black,inner sep=1.2pt},
  depot/.style={circle,draw,fill=white,inner sep=1.5pt},
  first/.style={draw=blue!70!black,line width=0.8pt,->},
  second/.style={draw=orange!85!black,line width=0.8pt,->}
]
  \draw[blue!70!black,line width=1.7pt] (0,0)--(1.6,0);
  \draw[black!65,line width=1.3pt] (1.6,0)--(2.4,0);
  \draw[orange!85!black,line width=1.7pt] (2.4,0)--(4,0);
  \foreach \x in {0,0.8,1.6,2.4,3.2,4} \node[unit] at (\x,0) {};
  \node[depot,label=above:$r_g$] (r) at (3.2,1.05) {};
  \node[depot,label=above:$h_g$] (h) at (4,1.05) {};
  \draw[black!65,line width=1.3pt] (h)--(r)--(3.2,0);
  \node at (2.87,0.65) {$e_P$};
  \draw[first] (3.07,0.96) -- (3.07,0.16) -- (0,0.16);
  \draw[first] (3.07,0.75) -- (3.07,0.50);
  \draw[second] (3.92,0.93) -- (3.33,0.93) -- (3.33,0.15) -- (4,0.15);
  \draw[second] (3.88,0.93) -- (3.53,0.93);
  \draw[second] (3.33,0.85) -- (3.33,0.45);
  \draw[second] (4,0.15) .. controls (4.28,0.15) and (4.28,-0.22) .. (4,-0.22) -- (2.4,-0.22);
  \node[below] at (0,-0.48) {$v(u_1)$};
  \node[below] at (1.6,-0.48) {$v(u_3)$};
  \node[below,orange!85!black] at (2.4,-0.48) {$P_2$};
  \node[below] at (3.2,-0.48) {$v(u_k)$};
  \node[below] at (4,-0.48) {$v(u_Q)$};
  \node[below,blue!70!black] at (0.8,-0.48) {$P_1$};
\end{tikzpicture}
\caption{$u_k\in P_2$}
\end{subfigure}
\caption{The walks for $P_1$ and $P_2$, with $Q=6$ and $b_{g,1}=3$.}
\label{fig:tree-construction}
\end{figure}

For a component $W$, if $B_W^0$ contains units of demand, we serve these units by traversing $\T_W$ twice from $r_W$ and shortcutting to $r_W$ and the clients whose demand is in $B_W^0$. This gives a tree of cost at most $2c(\T_W)$, summing to at most $2\sum_W c(\T_W)$ over all components. Including dummy units and $B_W^0$, each $r\in U^*$ is allocated $b_r$ units, so deleting dummy units leaves at most $b_r$ units served at $r$. Combining the bounds above with $\sum_W c(\T_W)\le C_Q(f)$ and $L(f)\le QC_Q(f)$, the total cost of all trees is at most
\[
 2\sum_P\bigl(c(P)+c(e_P)\bigr)+\sum_{g\text{ new}}c(r_g,h_g)+2\sum_W c(\T_W)
 \le 10C_Q(f).
\]

To bound the tree count at $r\in U^*$, note that its $b_r$ units form a consecutive interval of length $b_r$ in the label list, which intersects at most $\ceil{\frac{b_r}{Q}}+1$ blocks. If it intersects $\ceil{\frac{b_r}{Q}}+1$ blocks, its counts $\alpha_r,\beta_r$ in the first and last intersected blocks satisfy $\alpha_r+\beta_r\le Q$. Each intersected block yields at most one tree at $r$ with load at most $r$'s count in that block. If both the first and last intersected blocks produce nonempty trees at $r$, taking a spanning tree of their union preserves their served nodes, keeps the load at most $\alpha_r+\beta_r\le Q$, and does not increase cost. Thus at most $\ceil{\frac{b_r}{Q}}$ trees remain at $r$.

If $Q$ divides $b_r$ for every $r\in U^*$, there are no new labels or incomplete blocks. The unit $u_k$ used to attach $P$ has label $r_g$ and hence original depot $r_g$. Thus $v(u_k)$ is either $r_g$ (if $u_k$ is a dummy unit) or a client whose demand in $P$ is served by $r_g$. Shortcut $P$ to $\{v(u_k)\}$ and the clients whose demand is in $P$, add $e_P$, and take a spanning tree. Each tree then costs at most $c(P)+c_{P,g}$, reducing the total tree cost to at most $2\sum_W c(\T_W)+\frac{L(f)}{Q}\le 3C_Q(f)$.

\paragraph{Computing with binary counts.}
We store the at most $nm$ counts $x_{v,r}$ and $m$ dummy counts. Introducing new labels and setting aside incomplete blocks split $\O(m)$ counts. We keep each resulting count consecutive when ordering the units by location. Thus only $\O(nm)$ blocks contain units from different counts. Blocks containing units from only one count are assigned to their label, with identical blocks recorded by their multiplicity; minimum-cost flow assigns the other blocks. The tree merges add $\O(m)$ records, so the algorithm runs in polynomial time and outputs $\O(nm)$ tree records.
\end{proof}

\section*{Acknowledgement}
OpenAI's ChatGPT (GPT-6 Pro) was used for research exploration, proof checking, and manuscript preparation.
The authors independently verified the results, reorganized the paper's structure, rewrote the manuscript, and take full responsibility for its content.

\bibliographystyle{apalike}
\bibliography{HCCFL}
\end{document}